\documentclass{ieeeaccess}
\usepackage{cite}
\usepackage{amsmath,amssymb,amsfonts}
\usepackage{algorithmic}
\usepackage{graphicx}
\usepackage{booktabs}
\usepackage{multicol}
\usepackage{tabularx}
\usepackage{makecell}
\usepackage{textcomp}
\usepackage[edges]{forest}
\usepackage[hyphens]{url}
\usepackage[nolist]{acronym}
\usepackage[pagewise]{lineno}

\usepackage{pgfplots}
\usepackage{pgfplotstable}
\pgfplotsset{compat=1.7}
\usepackage{tikz}
\definecolor{accessblue}{cmyk}{1, 0.3, 0, 0.2}
\definecolor{greycolor}{cmyk}{0,0,0,.8}

\newcommand{\cmark}[1][]{\tikz[x=1em, y=1em]\fill[#1] (0,.35) -- (.25,0) -- (1,.7) -- (.25,.15) -- cycle;}
\newcommand{\xmark}[1][]{%
\tikz[x=1em, y=1em, line width = .15ex]{
    \draw[line cap=round, #1] (0,0) to[bend left=6] (0.45,0.45);
    \draw[line cap=round, #1] (0,0.45) to[bend right=2] (0.45,0);
    }
}

\NewSpotColorSpace{PANTONE}
\AddSpotColor{PANTONE} {PANTONE3015C} {PANTONE\SpotSpace 3015\SpotSpace C} {1 0.3 0 0.2}
\SetPageColorSpace{PANTONE}

\usepackage{bm}
\makeatletter
\AtBeginDocument{\DeclareMathVersion{bold}
\SetSymbolFont{operators}{bold}{T1}{times}{b}{n}
\SetSymbolFont{NewLetters}{bold}{T1}{times}{b}{it}
\SetMathAlphabet{\mathrm}{bold}{T1}{times}{b}{n}
\SetMathAlphabet{\mathit}{bold}{T1}{times}{b}{it}
\SetMathAlphabet{\mathbf}{bold}{T1}{times}{b}{n}
\SetMathAlphabet{\mathtt}{bold}{OT1}{pcr}{b}{n}
\SetSymbolFont{symbols}{bold}{OMS}{cmsy}{b}{n}
\renewcommand\boldmath{\@nomath\boldmath\mathversion{bold}}}
\makeatother

\def\BibTeX{{\rm B\kern-.05em{\sc i\kern-.025em b}\kern-.08em
    T\kern-.1667em\lower.7ex\hbox{E}\kern-.125emX}}

\begin{acronym}[AAAAAA]
    \acro{3G}{third generation}
    \acro{3GPP}{third generation partnership project}
    \acro{4G}{fourth generation}
    \acro{5G}{fifth generation}
    \acro{6G}{sixth generation}
    \acro{5GC}{5G core network}
    \acro{AAU}{active antenna unit}
    \acro{AI}{artificial intelligence}
    \acro{ARIMA}{autoregressive integrated moving average}
    \acro{ASIC}{application-specific integrated circuit}
    \acro{BBU}{baseband unit}
    \acro{BWP}{bandwidth part}
    \acro{BS}{base station}
    \acro{CPU}{central processing unit}
    \acro{COTS}{commercial off-the-shelf}
    \acro{CSIT}{channel state information at the transmitter}
    \acro{CSI}{channel state information}
    \acro{CSI-RS}{channel state information-reference signals}
    \acro{CU}{centralised unit}
    \acro{C-RAN}{cloud radio access network}
    \acro{DFE}{digital front-end}
    \acro{DL}{downlink}
    \acro{DRAM}{dynamic random access memory}
    \acro{DU}{distributed unit}
    \acro{DNN}{deep neural network}
    \acro{DQN}{deep Q-learning network}
    \acro{EC2}{Elastic Compute Cloud}
    \acro{EE}{energy efficiency}
    \acro{FLOPS}{floating point operations per second}
    \acro{FPGA}{field-programmable gate array}
    \acro{gNB}{next generation NodeB}
    \acro{GPU}{graphics processing unit}
    \acro{MCS}{modulation and coding scheme}
    \acro{MANO}{management and network orchestration}
    \acro{mMIMO}{massive multiple-input multiple-output}
    \acro{ML}{machine learning}
    \acro{MSR}{machine specific registers}
    \acro{NFVI}{network function virtualisation infrastructure}
    \acro{NIC}{network interface card}
    \acro{NR}{new radio}
    \acro{NG-RAN}{next-generation radio access network}
    \acro{O-RAN}{Open Radio Access Network}
    \acro{PA}{power amplifier}
    \acro{PDCP}{packet data convergence protocol}
    \acro{PHY}{physical layer}
    \acro{QoS}{quality of service}
    \acro{RAN}{radio access network}
    \acro{RAT}{radio access technology}
    \acro{RAPL}{running average power limit}
    \acro{RF}{radio frequency}
    \acro{RFE}{radio front-end}
    \acro{RRH}{remote radio head}
    \acro{RU}{radio unit}
    \acro{SARSA}{state–action–reward–state–action}
    \acro{SDAP}{service data adaptation protocol}
    \acro{SE}{spectral efficiency}
    \acro{TRP}{transmission-reception point}
    \acro{UE}{user equipment}
    \acro{VNF}{virtual network function}
    \acro{xHaul}{crosshaul transport network}
    \acro{PRB}{physical resource block}
    \acro{HARQ}{hybrid automatic-repeat-request}
    \acro{NN}{neural network}
    \acro{RM}{recommendation model}
    \acro{CV}{computer vision}
    \acro{NLP}{natural language processing}
    \acro{TN}{transport network}
    \acro{SCQP}{Successive convex quadratic programming}
    \acro{vBS}{virtualised base station}
    \acro{UL}{uplink}
    \acro{RRM}{radio resource management}
    \acro{CRM}{computational resource management}
    \acro{VM}{virtual machine}
    \acro{PDCP}{packet data convergence protocol}
    \acro{SDAP}{service data adaptation protocol}
    \acro{RFE}{radio front-end}
    \acro{MAC}{medium access control}
    \acro{RLC}{radio link control}
    \acro{RRC}{radio resource control}
    \acro{BB}{baseband}
    \acro{CPRI}{common public radio interface}
    \acro{PON}{passive optical network}
    \acro{FTTN}{fibre-to-the-node}
    \acro{PtP}{point-to-point}
    \acro{SM}{sleep mode}
    \acro{mmWave}{millimeter wave}
    \acro{SON}{self-organising network}
    \acro{RS}{rate splitting}
    \acro{D2D}{device-to-device}
    \acro{NOMA}{nonorthogonal multiple access}
    \acro{MIMO}{multiple-in multiple-out}
    \acro{MILP}{mixed-integer linear programming}
    \acro{LOS}{line-of-sight}
    \acro{DSL}{digital subscriber line}
    \acro{HFC}{hybrid fiber coaxial}
    \acro{SNR}{signal-to-noise-ratio}
    \acro{GPP}{general purpose processor}
    \acro{NFV}{network function virtualisation}
    \acro{OPEX}{operational expenditure}
    \acro{LLM}{large-language model}
    \acro{DRL}{deep reinforcement learning}
    \acro{MCPA}{multicarrier power amplification}
    \acro{ANN}{artificial neural network}
    \acro{NTN}{nonterrestrial network}
    \acro{OWC}{optical wireless communication}
    \acro{THz}{terahertz}
    \acro{SINR}{signal-to-noise-plus-interference ratio}
    \acro{TRX}{transceiver chain}
    \acro{IRS}{intelligent reflective surfaces}
    \acro{SDR}{software-defined radio}
    \acro{ITU}{International Telecommunications Union}
    \acro{FDD}{frequency division duplexing}
    \acro{TDD}{time division duplexing}
    \acro{PSU}{power supply unit}
    \acro{NG}{next-generation}
    \acro{Xn}{cross-node}
    \acro{SOM}{service management and orchestration}
\end{acronym}

\begin{document}
\history{Date of publication xxxx 00, 0000, date of current version xxxx 00, 0000.}
\doi{10.1109/ACCESS.2024.0429000}

\title{A Survey on AI-driven Energy Optimisation in Terrestrial Next Generation Radio Access Networks}
\author{\uppercase{Kishan Sthankiya}\authorrefmark{1}, \IEEEmembership{Student Member, IEEE},
\uppercase{Nagham Saeed}\authorrefmark{2}, \IEEEmembership{Senior Member, IEEE},
\uppercase{Greg McSorley}\authorrefmark{3}
\uppercase{Mona Jaber}\authorrefmark{1}, \IEEEmembership{Member, IEEE}
and Richard G. Clegg,
\authorrefmark{1},
}

\address[1]{School of Electronic Engineering and Computer Science, Queen Mary University of London, London, E1 4NS, U.K.}
\address[2]{School of Computing and Engineering, University of West London, London W5 5RF, U.K.}
\address[3]{Applied Research BT, IP5 3RE Suffolk, U.K.}
\tfootnote{This research was supported by the University of West London Knowledge Exchange Seed Fund under Project SF16 – 15036 “Working towards Energy Efficient wireless network: A collaboration with BT Group to study and evaluate the AI and ML usage” and the UK Engineering and Physical Sciences Research Council (EPSRC) grant EP/V519935/1.}

\markboth
{Sthankiya \headeretal: A Survey on AI-driven Energy Optimisation in Terrestrial NG-RAN}
{Sthankiya \headeretal: A Survey on AI-driven Energy Optimisation in Terrestrial NG-RAN}

\corresp{Corresponding author: Nagham Saeed (e-mail: nagham.saeed@uwl.ac.uk).}

\begin{abstract}
This survey uncovers the tension between AI techniques designed for energy saving in mobile networks and the energy demands those same techniques create. 
We compare modeling approaches that estimate power usage cost of current commercial terrestrial next-generation radio access network deployments. 
We then categorize emerging methods for reducing power usage by domain: time, frequency, power, and spatial. 
Next, we conduct a timely review of studies that attempt to estimate the power usage of the AI techniques themselves. 
We identify several gaps in the literature.
Notably, real-world data for the power consumption is difficult to source due to commercial sensitivity.
Comparing methods to reduce energy consumption is beyond challenging because of the diversity of system models and metrics.
Crucially, the energy cost of AI techniques is often overlooked, though some studies provide estimates of algorithmic complexity or run-time. 
We find that extracting even rough estimates of the operational energy cost of AI models and data processing pipelines is complex. Overall, we find the current literature hinders a meaningful comparison between the energy savings from AI techniques and their associated energy costs. Finally, we discuss future research opportunities to uncover the utility of AI for energy saving.
\end{abstract}


\titlepgskip=-21pt

\maketitle
\section{Introduction}
\label{sec:introduction}
\PARstart{E}{nergy} and carbon reductions for mobile networks have never been more important given the goal to meet net-zero by 2050 and user data traffic is estimated to rise five-fold in moving to \ac{5G}. The \ac{RAN} remains a significant energy consumer (estimated 87\% of network operations and up to 40\% of \ac{OPEX})~\cite{gsma_mobile_2023}. This has led to a push for \ac{AI} driven solutions for energy reduction in \ac{RAN} deployments~\cite{soldati_approaching_2023}. However, \ac{AI} itself can have a large energy cost.  Estimates for the energy cost of training a \ac{LLM} such as OpenAI's GPT-3 stand at 1,287 MWh, whereas estimates for operational energy demand stand at 564 MWh~\cite{de_vries_growing_2023}. Meta~\cite{wu_sustainable_2022} estimates the energy footprint of \ac{AI} inference of an in-house \ac{RM} to account for 40\% of the whole model energy consumption. Similarly, Google~\cite{patterson_carbon_2022} estimates \ac{AI} inference alone accounted for 9\% of their total energy use between 2019 and 2021.

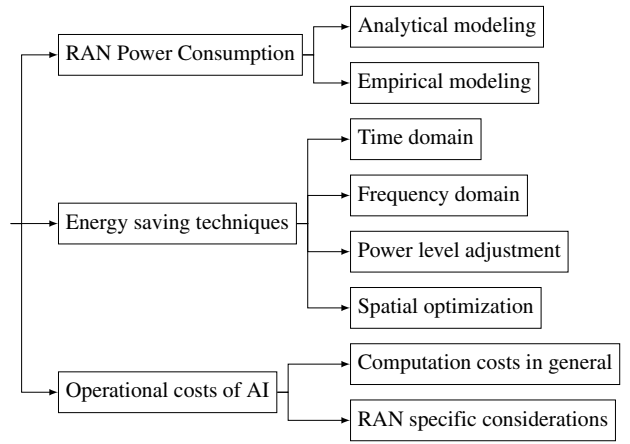
\begin{figure}[ht]
    \centering
    \resizebox{3.3in}{!}{
    \begin{forest}
        for tree={
            grow'=east,        
            forked edges,
            draw, align=center,
            edge={-latex},
            anchor=west,
            child anchor=west,
            l sep+=10pt,        
            tier/.wrap pgfmath arg={tier #1}{level()}
        }
        [ ,draw=none, edge=none
            [RAN Power Consumption
                [Analytical modeling]
                [Empirical modeling]
            ]
            [Energy saving techniques
                [Time domain]
                [Frequency domain]
                [Power level adjustment]
                [Spatial optimization]
            ]
            [Operational costs of AI
                [Computation costs in general]
                [RAN specific considerations]
            ]
        ]
    \end{forest}
    }
    \caption{High-level taxonomy of topics covered in this survey}
    \label{fig:paper_structure}
\end{figure}

This survey paper focuses on the \ac{RAN} and looks at how \ac{AI}/\ac{ML} can be used to reduce power consumption but also to consider the power consumption of the required \ac{AI} inference. In particular, we investigate if the power cost of algorithms to reduce energy consumption can ever approach or exceed the energy saved. A high-level overview of the topics covered can be found in Fig.~\ref{fig:paper_structure}. We begin with a survey of \ac{RAN} power consumption models asking whether the research community has a good and well-evidenced model of the power used by a \ac{RAN} and this will be the basis for an accurate estimate of power saved. Following this we look at the different optimization models used to reduce power consumption considering the physical techniques used (what \ac{RAN} parameters are being changed to get the power savings) and what \ac{AI} techniques are being used to achieve this. We limit our survey to techniques that are already deployed or standardized and ready to deploy \ac{RAN} technologies and report results with improvements in energy saving or energy efficiency. Finally, we investigate the question of how much energy might be consumed by \ac{AI} models deployed for energy reduction. Because timeliness is vital in a rapidly moving field like this one we have chosen papers published in 2020 or afterwards with a few exceptions where older papers are a vital part of later understanding.

\label{sec:intro-paper-structure}
To answer the questions above, this survey is structured as follows.
The remainder of this section reviews related survey papers highlighting the key differences of this work. This is followed by an outline of the scope of this survey. 
Section~\ref{sec:ng-ran-archi} introduces the 5G \ac{RAN} architecture as a grounding for discussing power models in Section~\ref{sec:power-models}.
In Section~\ref{sec:energy-saving}, we survey the literature on energy-saving techniques, highlighting the key contributions in the time, frequency, power and spatial domains. In Section~\ref{sec:ai-ml-energy-ops}, we review the areas that impact the energy cost of \ac{AI} inference in the \ac{NG-RAN} and, where required, draw in the broader research literature. Finally, in Section~\ref{sec:conclusions}, we present concluding remarks with suggestions for future research directions. 

A note on terminology: the terms \ac{ML} and \ac{AI} are often used somewhat interchangeably, to avoid the somewhat clumsy \ac{ML}/\ac{AI} we will use \ac{AI} throughout in this survey unless there is a good reason to prefer the term \ac{ML} in context (for example where the authors of a paper use this term). Many (but not all) techniques discussed have both a training phase (done once only or at infrequent intervals) which produces the parameters used by the model and an inference phase that produces the answer given a set of parameters. The training phase is typically more computationally intensive but, in a production network, the inference phase needs to be used every time an answer is required hence cannot be avoided as an operational cost. 

\subsection{Related Survey Papers}
\begin{table*}[!htbp]
    \centering
    \caption{A comparison of our work with other survey papers since 2020}
    \label{tab:surveys}
    \begin{tabular}{|m{3.5cm} | m{0.6cm} |m{0.6cm} | m{0.6cm} | m{0.6cm} | m{0.6cm} |m{0.6cm} |m{1.2cm} | } \hline
    & \multicolumn{7}{c|}{\bf Refs} \\ \cline{2-8}
    {\bf Topic} & {\cite{rodoshi_resource_2020}} & 
    {\cite{lopez-perez_survey_2022}} & 
    {\cite{brik_deep_2022}} & 
    {\cite{abubakar_energy_2023}} & 
    {\cite{larsen_toward_2023}} &
    {\cite{mao_ai_2022}} &
    {\bf Our work}    \\ \hline 

    Current RAN challenges & 
    \makecell{\xmark} &
    \makecell{\cmark} &
    \makecell{\cmark} &
    \makecell{\cmark} &
    \makecell{\xmark} &
    \makecell{\xmark} &
    \makecell{\cmark} \\ \hline

    Empirical RAN power models &
    \makecell{\xmark} &
    \makecell{\cmark} &
    \makecell{\xmark} &
    \makecell{\cmark} &
    \makecell{\xmark} &
    \makecell{\cmark} &
    \makecell{\cmark} \\ \hline

    AI power factors &
    \makecell{\xmark} &
    \makecell{\xmark} &
    \makecell{\xmark} &
    \makecell{\cmark} &
    \makecell{\xmark} &
    \makecell{\cmark} &
    \makecell{\cmark} \\ \hline

    Sleep modes &
    \makecell{\cmark} &
    \makecell{\cmark} &
    \makecell{\cmark} &
    \makecell{\cmark} &
    \makecell{\cmark} &
    \makecell{\cmark} &
    \makecell{\cmark} \\ \hline

    Rate splitting &
    \makecell{\xmark} &
    \makecell{\xmark} &
    \makecell{\xmark} &
    \makecell{\xmark} &
    \makecell{\cmark} &
    \makecell{\xmark} &
    \makecell{\cmark} \\ \hline

    Interference management &
    \makecell{\cmark} &
    \makecell{\xmark} &
    \makecell{\xmark} &
    \makecell{\xmark} &
    \makecell{\cmark} &
    \makecell{\cmark} &
    \makecell{\cmark} \\ \hline

    \end{tabular}
    \normalsize
\end{table*}

Reviews that focus on \ac{AI} for power-saving in the \ac{RAN} are well studied and the major competing surveys in this space since 2020 are~\cite{rodoshi_resource_2020,lopez-perez_survey_2022,brik_deep_2022,abubakar_energy_2023,larsen_toward_2023,mao_ai_2022}. To the best of the authors knowledge, the novelty in this work is an emphasis on also considering the energy cost of \ac{AI}. A summary of the other papers compared with this one is given in Table \ref{tab:surveys}. Some surveys prefer to give their attention to future enablers for 6G technology~\cite{rodoshi_resource_2020,larsen_toward_2023,mao_ai_2022} that are not covered by this paper. By contrast, our focus is on technologies deployed today or standardized and ready for deployment,
by studies where quantitative energy savings are reported which could be immediately beneficial.

While three of the other studies include power consumption models~\cite{lopez-perez_survey_2022, abubakar_energy_2023, mao_ai_2022} the surveys~\cite{lopez-perez_survey_2022, abubakar_energy_2023} do not break down models into analytical or empirical and the other~\cite{mao_ai_2022} uses only older \ac{3G} models of power. This survey, by contrast, offers a timely breakdown of the analytical and empirical power consumption models using current generation technology. Most surveys do not cover the downside of optimization, the energy cost of \ac{AI}. While~\cite{abubakar_energy_2023} highlights computational effort as the number of operations per second, this still misses a huge number of factors that contribute to algorithmic power consumption. By contrast, this survey details the factors involved in the power consumption of an \ac{AI} algorithm. The only survey the authors found that covers this field reasonably is~\cite{garcia-martin_estimation_2019} but this survey is now five years old whereas we focus on  \ac{AI} techniques from 2020 onward. This is the key differentiator between this survey and others in the field.

Other surveys have included a number of works that look at techniques to manage energy consumption in the \ac{RAN} but we believe this to be the most up-to-date and complete. Both~\cite{lopez-perez_survey_2022,larsen_toward_2023} give extensive explanations on how sleep modes and different levels of shutdowns work for power saving at a base station, whereas~\cite{mao_ai_2022} focus on ways to maximize sleep duration. These surveys are from 2022 and 2023 respectively so our survey complements and updates them. 

Interference management for energy efficiency is covered in~\cite{rodoshi_resource_2020, mao_ai_2022} but the former focuses only on remote radio head clustering in cloud RAN and the latter on only techniques that modify transmit power. 
The survey~\cite{larsen_toward_2023} highlights the novelty of \ac{RS} for efficiency which we also cover. 
In this survey, we look at how scheduling techniques can help to reduce delay, power consumption and maximize profit for an operator. This is a promising area of research, but discussions in the literature have been sparse in recent works. For instance,~\cite{lopez-perez_survey_2022} do not consider it and~\cite{abubakar_energy_2023} limits their discussion to one study. In contrast,~\cite{larsen_toward_2023} covers multiple operator sharing and baseband workload scheduling. 

\subsection{Scope and Contributions}
This survey focuses on the impact of \ac{AI}-based algorithms on reducing power usage and the energy cost of that \ac{AI} and focuses on developments since 2020 (although older papers are included, particularly in considering power estimation, where they remain the state-of-the-art). The survey is of viable techniques used in current 5G installations where energy savings are explicitly reported. We categorically do not cover supply-side power management technologies such as improved power generation, renewable energy, battery, or smart grid technologies. We recognize the potential utility of post \ac{5G} technologies, such as \acp{NTN}~\cite{azariEvolutionNTN2022}, \ac{OWC}~\cite{palitharathnaSLIVER2023, papanikolaouSimultaneous2024} and \ac{THz}~\cite{wu_lidar-aided_2022, rahimMultiIRSAidedTerahertzNetworks2024} communications but these are not our focus here. 
Our main contributions include:
\begin{enumerate}[\IEEEsetlabelwidth{1}]
    \item Identifying the analytical and empirical power consumption models in the \ac{RAN}. We compare how power consumption models are delineated based on the scope and architecture.
    \item A timely review of leading research on \ac{RAN} \ac{EE}, classifying the studies by their leading degree of freedom (e.g.\ time, frequency, power, and spatial domains).
    \item Discussion of the factors that impact the operational energy cost of using \ac{AI} techniques as this may mitigate any savings made.
\end{enumerate}
Following the survey, we highlight the gaps in the existing research, providing insights into directions for future research on \ac{AI} for improving \ac{RAN} \ac{EE}.

\section{NG-RAN Architecture}
\label{sec:ng-ran-archi}
When planning a RAN deployment, the design is typically over-provisioned in order to be future-proof (because deployment is costly), and to cope with peak load. 
Breaking down hardware functions into decoupled logical units creates opportunities for more granular scaling, gains in power consumption and efficiency.
Hosting network functions in different physical locations and hardware can allow efficient responses to changes in demand patterns.
In this section, our focus is to describe the logical units that constitute the \ac{NG-RAN}, forming a foundation for later discussions on similarities and differences between power models.
\begin{figure}
    \centering
    \includegraphics[width=\columnwidth]{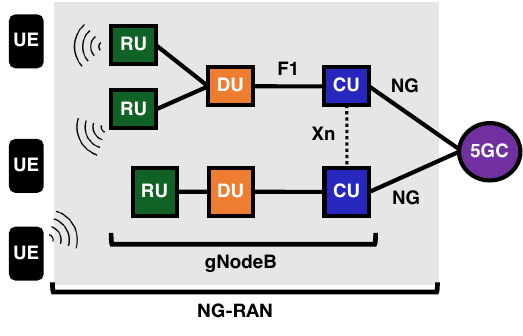}
    \caption{Overview of \ac{5G} System. Composed of UE, NG-RAN (shaded) and 5G Core Network.}
    \label{fig:5GS-Overview}
\end{figure}
Depicted in Fig.~\ref{fig:5GS-Overview}, the \ac{NG-RAN} is a collection of several of base station known as \ac{gNB}~\cite{3gpp_NG-RAN_Overall_Desc_ts38_300_v17_6_0}. Each \ac{gNB} contains one \ac{CU} and one or more \acp{DU}\footnote{Formally, a CU and DU are referred to as gNB-CU and gNB-DU, respectively, but we omit the prefixes for simplicity.}.
Complementary to the \ac{3GPP} specifications, the \ac{O-RAN} Alliance defines standards to promote architectures that use open interfaces while fostering hardware disaggregation, flexibility and network intelligence~\cite{brik_deep_2022}. This makes \ac{O-RAN} a key technology to allow \ac{AI} to be used in \ac{RAN}. In addition to the aforementioned logical units, the \ac{O-RAN} describes \ac{RU}\footnote{The 3GPP specifications do not formally include \ac{RU} as part of the \ac{gNB}. However, we do here because of the significant impact on power consumption.}, where each \ac{DU} connects to one or more \acp{RU}. The connections between physical or logical nodes in the \ac{3GPP} specifications for \ac{NG-RAN} are described in~\cite{3gpp_arch_desc_ts_38.401_v17.6.0}. The edges between logical units describe the \ac{xHaul}. 

The radio signals between \ac{UE} and \ac{NG-RAN} are transmitted and received by the \ac{RU}. The \ac{RU} are always located at network operator cell sites, which are spatially distributed to ensure geographical coverage. The \ac{RU} converts between the analogue radio signals used by antennae and the digital signals used by the \ac{DU}. The \ac{DU} connects to one or more \acp{RU} and the \ac{CU} typically handles the higher level protocol stack. 

As previously mentioned, the \ac{xHaul} describes the transport network supporting the sending and receiving of signals between \ac{RU}, \ac{DU} and \ac{CU} nodes. It is made up of the fronthaul (\ac{RU}-DU), midhaul (DU-CU) and backhaul (CU-Core).

\section{RAN Power Consumption Modeling}
\label{sec:power-models}
In order to properly evaluate the influence of \ac{AI} on the \ac{EE} of the \ac{NG-RAN}, it is crucial to understand the assumptions of models that provide estimates of power consumption. These models must consider the enabling technologies used in the \ac{NG-RAN} while remaining flexible to evolving RAN architectures. Two main types of studies emerge from the literature, \emph{analytical models} and \emph{empirical models}. Analytical models here attempt to derive equations from physical principles that could estimate power consumption given correct input data and physical parameters. By contrast, the empirical models use measured data to attempt to ground these estimates in the real world. In this section, we present prominent models and approaches used for \ac{NG-RAN} power consumption. It should be noted that publications creating new power models are far less frequently published and, hence, the references in this section are older since in this area papers that represent the current best state of understanding can be more than ten years old. 

\begin{figure*}[htbp!]
    \centering
    \includegraphics[width=\textwidth]{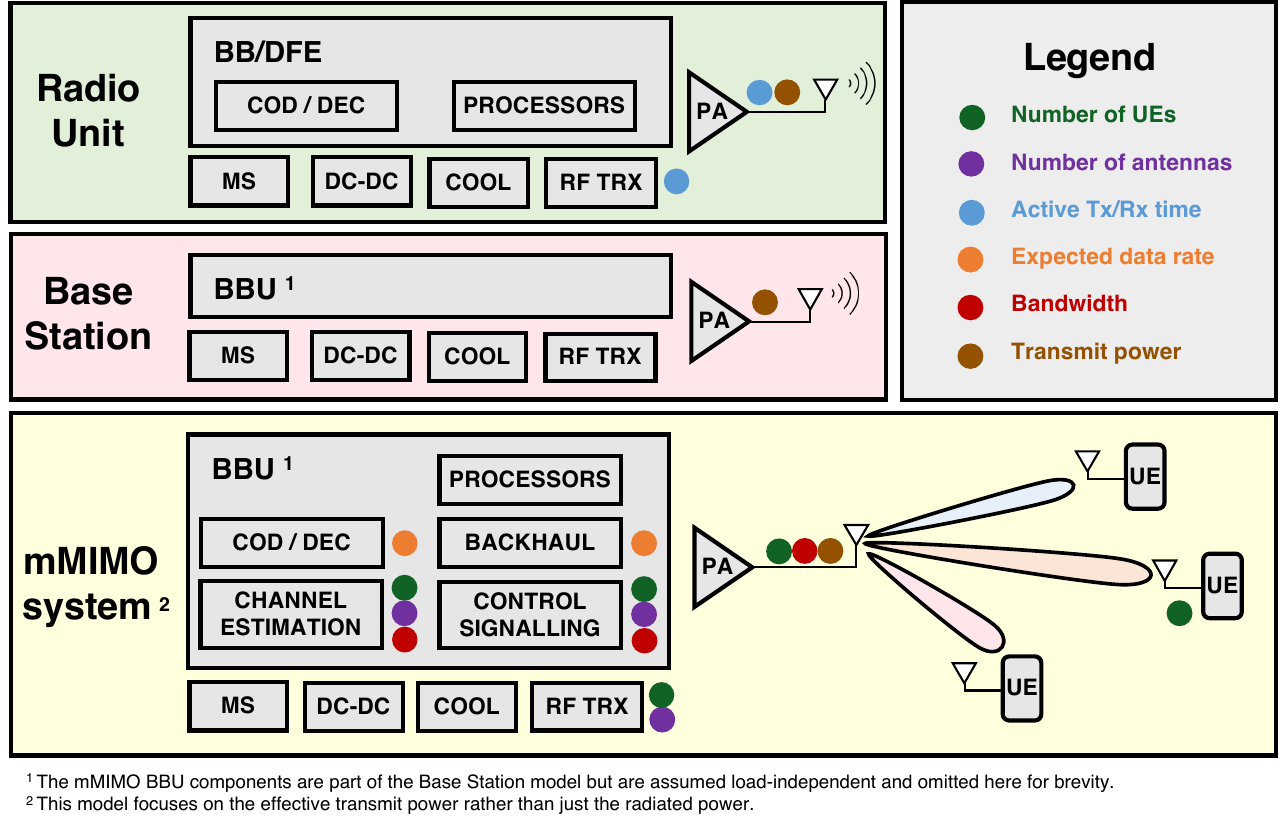}
    \caption{A comparison of power consumption models from the literature focused on the radio unit, base station and Massive Multiple-Input and Multiple-Output (mMIMO) system. Components include the baseband-digital front end (BB/DFE), baseband unit (BBU), channel coding and decoding (COD/DEC), mains power supply losses (MS), direct current conversion losses (DC-DC), active cooling losses (COOL), radio frequency transceiver (RF TRX), power amplifier (PA) and user equipment (UE).
    Grey components without a dot indicate a component with load-independent power consumption. Components with a dot represent a dynamic power consumption, where the color represents the influencing factor. 
    }
    \label{fig:power-models}
\end{figure*}

\textcolor{black}{A functional split describes the division of the baseband processing chain and which logical nodes are responsible for carrying out that function.}
When considering the power consumption of the logical nodes in the \ac{NG-RAN} (e.g. \ac{RU}, \ac{DU}, or \ac{CU}), it is essential to consider how power consumption will be affected by the chosen functional split since the network functions hosted at each node will impact the computational load and therefore the energy consumed. 
Some authors have looked in detail at the effects of functional split on energy efficiency~\cite{arzo_study_2020, gkatzios_optimized_2020, moreira_zorello_power-efficient_2022, amiri_energy-aware_2023}. For example, \textit{service differentiation}, a technique using backup \acp{VNF} to improve resilience and \acp{CPU} over-provisioning to decrease the queuing delay of the \acp{VNF}, improves \ac{EE}~\cite{arzo_study_2020}. Live migration of virtualized resources reduces the number of ``switched-on'' servers reducing the average energy consumption by 8\%~\cite{gkatzios_optimized_2020}. Similarly, placement of \acp{CU} and \acp{DU} in a metro access network is solved using a heuristic~\cite{moreira_zorello_power-efficient_2022} saving almost 8\% of total power when compared to a static \ac{MILP} approach. In contrast,~\cite{amiri_energy-aware_2023} use \ac{DRL} to achieve dynamic \ac{VNF} splitting in an \ac{O-RAN} scenario by as much as 63\% compared to a Greedy algorithm approach. Therefore, it is crucial to understand the power consumption of each individual component, including the \ac{RU}, supporting infrastructure for both distributed and centralized units (whether physical or virtualized), and the \ac{xHaul}. A recent report by NGMN for Green Future Network\footnote{https://www.ngmn.org/} emphasizes the importance of hardware metering when standard COTS (Commercial-off-the-shelf) that would host some of these \ac{VNF} with the aim of determining the energy consumption analyses how the cloud model could be harnessed to optimize the energy efficiency.

\subsection{Analytical Models}
Analytical models attempt to construct estimates for power consumption from equations based on physical principles. Our survey found three major power models for \ac{RAN} networks. However, each covers a slightly different part of the system, and each makes different assumptions about which components have constant power consumption and which components vary with load. Fig.~\ref{fig:power-models} will be used to illustrate which components are included in each of the three major models we cover. The top (green) box represents the \ac{RU} model from~\cite{wesemann_energy_2023}, the middle (pink) box represents the base station \ac{BS} model known as EARTH~\cite{auer_how_2011} and the bottom (yellow) box represents the mMIMO model from~\cite{bjornson_massive_2017}. The models in these works are extremely detailed and here a high-level view is given. 

The authors in~\cite{wesemann_energy_2023} formulate a power model for an \ac{RU} (green box in Fig.~\ref{fig:5GS-Overview}). This model accounts for current (3GPP {Release-18}) and future \ac{MIMO} architecture considering the scaling effects on power consumption of discontinuous transmission and reception schemes, antenna muting and chip processing in addition to radiated transmit power. The mapping of components of this \ac{RU} model can be found in Fig.~\ref{fig:power-models} (green shaded box). The \ac{RU} power consumption model~\cite[Eq.1]{wesemann_energy_2023} is presented as\footnote{The original paper gives a more detailed equation, whereas we summarize terms here.}:

\begin{align}
    \label{eq:ru}
    P_\text{RU} = \beta M P_\text{Tx}^\text{dyn} +  (1-\beta) M P_\text{Rx}^\text{dyn} + M C P_\text{static}
\end{align}

where $M$ is the number of transceiver chains, $\beta$ is the uplink-to-downlink ratio for \ac{TDD}, $P_\text{Tx}^\text{dyn}$ is the dynamic transmit power, $P_\text{Rx}^\text{dyn}$ is the dynamic receive power, $C$ is the ratio of active computation power to total computation power, and $P_\text{static}$ accounts for the load-independent power consumption of the \ac{DFE} and \ac{BB}. 

As noted in~\cite{wesemann_energy_2023}, the burden of processing functions for \ac{BB} and \ac{DFE} are moving towards an integrated \ac{RU}, where it was once reserved for dedicated hardware called the \ac{BBU}. The combination of \ac{RU} + \ac{BBU}, commonly referred to as a \ac{BS}, is equivalent to the functions carried out in all three parts of a \ac{gNB}, namely the RU, DU and CU, as seen in Fig.~\ref{fig:5GS-Overview}. In lieu of reference models for \ac{DU} and \ac{CU} power consumption, traditional distributed \acp{RAN} power consumption models are commonplace. A cornerstone model, capturing more functions than the aforementioned \ac{RU} model is the EARTH framework~\cite{auer_how_2011} which maps the \ac{RF} output power ($P_\text{out}$), measured at the antenna interface, to the total supply power of a \ac{BS}. A visual representation of this \emph{Base Station} model is shown (pink shaded box) in Fig.~\ref{fig:power-models}. 
Abstracting away physical hardware components (e.g.~\ac{BBU}, \ac{PA}) and conversion losses (e.g.~cooling, mains supply) from their complex model, the power consumption for a \ac{4G} \ac{BS} in~\cite[Eq.~1]{auer_how_2011} is presented as:
\begin{align}
    P_\text{BS} =
    \begin{cases}
        M \cdot (P_\text{0} + \Delta_{ld} P_\text{out}), &0 < P_\text{out}\leqslant P_\text{max} \\
        M \cdot P_\text{sleep}, &P_\text{out} = 0
    \end{cases}
    \label{eq:auer}
\end{align}

where $M$ is the total number of BS antennas, $P_\text{0}$ is power consumption independent of RF output, $\Delta_{ld}$ is the gradient of the power consumption dependent on RF output power, $P_\text{out}$ is RF output power and $P_\text{sleep}$ is the sleep mode power consumption. 
It should be highlighted that when comparing \eqref{eq:ru} and \eqref{eq:auer}, $P_0 \neq P_\text{static}$, as the assumptions of what constitutes a load-independent factor of the power models differ, as illustrated in Fig.~\ref{fig:power-models}. 

Building on~\cite{auer_how_2011}, the authors in ~\cite{holtkamp_parameterized_2013} developed a tractable power model by factoring in \ac{PA} output range and transmission bandwidth. The \emph{GreenTouch} framework in~\cite{desset_towards_2013} and~\cite{b_debaillie_flexible_2015} further adopts a five-layer approach towards flexibility for future enabling technologies. The opportunities to reduce power consumption within this modeling approach suggest three strategies for reducing power consumption, such as reducing the RF output power, reducing load-independent power consumption and maximizing $P_\text{sleep}$, which we cover in Section~\ref{sec:energy-saving}.

Technological enablers for \ac{NG-RAN}, such as \ac{mMIMO} and \ac{NFV}, challenge assumptions for power consumption of past models. For example, the authors in~\cite{bjornson_optimal_2015} highlight the need for more sophisticated models when considering \ac{mMIMO} systems which increase the complexity of the \ac{BS} model. In particular, they assert that power consumption within the 
\acp{BBU}, \ac{RF} \acp{TRX} and \acp{PA} varies with the number of antennas and \acp{UE}. 

In~\cite{bjornson_massive_2017}, the authors derive the circuit power ($P_\text{CP}$) of a \ac{mMIMO} from the number of antennas $M$, number of users $K$, effective transmit power $P_\text{out}$ and gross rate $\Bar{R}$, for different linear processing schemes. A high-level view of the \ac{mMIMO} system power consumption model from~\cite[Eq.~21]{bjornson_optimal_2015} may be summarized as,
\begin{align} 
    P_\text{sys}^\text{mMIMO} = P_\text{out} + P_\text{CP}\left(M, K, \Bar{R}\right),    
\end{align}
The power consumption of hardware supporting the signaling between nodes is dependent on technology such as microwave radio, \ac{PON} and Ethernet~\cite{abubakar_energy_2023} in addition to factors such as network topology, capacity and activity. In~\cite{eramo_trade-off_2016}, the authors present an analytical model for the power consumption of the \ac{xHaul} as the sum of 1) power consumption as a function of the bandwidth of the \ac{CPRI} and Ethernet circuits between an access site and the central office; 2) power consumption of the radio stations and servers.
In~\cite{larsen_ran_2022}, the authors study how increasing \ac{RAN} coverage and capacity affects the \ac{xHaul} power consumption and which \ac{xHaul} parameters impact \ac{RAN} power consumption. Specifically, they model \emph{power consumption} of a \ac{xHaul} switch as:
\begin{align}
    \text{PC} 
    =
    P_\text{standby} + (P_\text{bit} \cdot N_{\text{bits}}) + (P_\text{pk} \cdot N_{\text{packets}}),
\label{eq:larsen-xhaul-model}
\end{align}

where $P_\text{standby}$ is the power consumption of a switch on standby, $P_\text{bit}$ is power consumption per bit traversing the switch, $N_\text{bits}$ is the number of bits, $P_\text{pk}$ is the power consumption of per packet and $N_\text{packets}$ is the number of packets. While the power consumption of the \ac{xHaul} is not the main focus of this review, we recognize the multiplicative effect of data volume on the power consumption of the transport network and on the wisdom of efforts towards more efficient data transmission and scheduling approaches, as discussed later in Section~\ref{sec:energy-saving}. 

\subsection{Empirical Models}
Empirical models are data driven and attempt to estimate power consumption from measurements of the system. We encountered two types of empirical model in this survey. The first type uses power consumption data from product information sheets provided by manufacturers enriched with features from traffic profiles or mobility. The second type conducts studies on testbeds to examine proposed designs and presents results from experimental methods, such as measurements from probes or power meters.

In~\cite{capone_modeling_2017} regression is used to analyze energy consumption data drawn from energy consumption sensors in 3G and 4G network deployments across 60 sites in three countries. They conclude to a good approximation a linear model relates traffic volume and emitted power and higher-order models lead to over-fitting.
More recently~\cite{piovesan_machine_2022} develop a power model for \ac{5G} multicarrier \ac{mMIMO} \acp{AAU}, where a single power amplifier can support multiple carriers using \ac{MCPA} technology and deep dormancy or symbol, channel or carrier level shutdown. They initially explore a data-centric approach using an \ac{ANN} and derive an analytical power consumption model based on the results. Compared to power models that do not account for MCPA effects \cite{bjornson_optimal_2015}, the proposed analytical model is described as being 1.5 times more accurate while maintaining a low mean absolute error of 5.6\% compared to the \ac{ANN} model.

The data from~\cite{capone_modeling_2017} is also the basis for~\cite{huttunen_base_2023}, which presents field measurements on data and visitor volumes. Combining these with parameters for different RATs (Radio Access Technologies), including \ac{5G} \acp{RU} from Nokia product data sheets, they calculate and extrapolate the base station power consumption in dense urban and suburban areas of Finland. Compared with a measurement campaign of the same base stations, the proposed theoretical model for 5G is better at predicting energy consumption in the dense urban area, with the caveat that there are more users of the same type in that area.

The discussed models lack good open 5G data. They use 4G data~\cite{capone_modeling_2017}, normalize the power consumption values~\cite{piovesan_machine_2022} (to maintain commercial security) or speculate on the power dynamics of 5G hardware based on manufacturer reported spectral efficiency~\cite{huttunen_base_2023}. This demonstrates a lack of open research with clear reporting of empirical power consumption within real networks. As an alternative approach, the gray literature (\emph{outside of formal commercial or academic publication}) can provide an intuition of peak power consumption of different types of \ac{5G} base stations. For example, actual power consumption from an anonymous operator shows that a 5G \ac{BS} under full load consumes approximately 1.4 kW~\cite{dappworks_front_2020} for vendor equipment supporting one band, whereas another source reports 4.7 kW~\cite{chen_dongxu_5g_2020} for a different vendor supporting 3--4 bands. To put this into context, the power consumption of a consumer workstation PC from a leading vendor~\cite{gartner_gartner_2023} ranges from 170 -- 300 W~\cite{lenovo_thinkstation_2023}, which is comparable to the reported \ac{BBU} power in~\cite{dappworks_front_2020, chen_dongxu_5g_2020}.  

As \ac{BBU} processing becomes disaggregated and workloads delegated to a \ac{DU} or \ac{CU}, these new nodes must cater for future growth. It is unsurprising, then, that datasheets for commercial servers advertised as suitable for \ac{DU} workloads report peak power consumption between 300--1800 W~\cite{hpe_hpe_2024, supermicro_5g_2024, rodriguez_dell_2023}. The EARTH power consumption model~\cite{auer_how_2011} is popular but predates the move toward \acp{vBS} and may not capture the power implications. Motivated by this, the authors in~\cite{ayala-romero_experimental_2021} approximate \ac{vBS} power consumption derived from experimental results from uplink transmissions in a testbed. 
Virtualization tackles the problem of over-provisioning, allowing resources to be scaled to the user demand and afford resilience. When considering \ac{vBS} it is important to know the cost of virtualization.

In~\cite{salvat_open_2023} the authors provide three open datasets, including the energy consumption of a \ac{vBS} as a function of a range of parameters within an \ac{O-RAN} compliant testbed. Similarly, the authors in~\cite{katsaros_power_2022} measure the power consumption (wattage) of software implemented \ac{PHY} for \ac{5G} NR using Intel's \ac{RAPL} \ac{MSR}, they measure \ac{CPU} and \ac{DRAM} power, estimating the measurement overhead as $\leqslant$1\%. This proves a valuable study to dimension the energy consumption of a software-defined \ac{NG-RAN} approach.

When \ac{NG-RAN} functions are virtualized it is important they still satisfy latency constraints. This is mitigated by having a dynamic functional split~\cite{matoussi_5g_2020}, which allows baseband processing to move closer to the cell site when required. Measuring the impact of different functional splits, the authors in~\cite{pawar_understanding_2020}~profile the energy consumption in an Open Air Interface testbed. Specifically, they attempt to profile the power consumption of a \ac{DU} and \ac{CU} by varying the \ac{CPU} clock frequency and channel bandwidth. They find cases where \ac{CPU} clock frequency could be reduced for use cases where full-buffer traffic is employed. These results are based on a single user and connecting to a \ac{RU} modeled using \ac{SDR}. Therefore, the applicability of these results when scaled up to operational network volumes and optimized \ac{C-RAN} datacenters, remains an unanswered question. Moreover, since the processing is decoupled from hardware (which vary between architectures), an extension of this study to quantify the computational complexity, per layer of the radio stack in the \ac{DU} and \ac{CU}, as a function of throughput, would provide a useful future-proof contribution. For example, what would be the empirical computational load (in \ac{FLOPS}) to run \ac{RLC} processing while ensuring a data rate of 100~Mbps?

Testbeds also provide a way to measure the energy consumption of the \ac{xHaul}. Considering the power consumption of access networks, the study~\cite{baliga_energy_2011} presents energy consumption figures for \ac{DSL}, \ac{HFC} networks, \ac{PON}, \ac{FTTN} and \ac{PtP} optical. The study found that optical networks are the most energy-efficient. Later studies by~\cite{granell_energy_2012} show that energy consumption does not grow proportionally with the number of ports, and~\cite{vishwanath_member_modeling_2014} show that high-capacity routers and switches use 80-90\% of their maximum power whilst at idle load. More recently, in~\cite{francini_low-cost_2015}, the authors provide a measurement methodology for power profiling based on a linear model for rate adaptation. They provide testbed measurements for two types of 24-port 1GbE switches (one with fixed ports, another with modular) and three routers (edge, fixed aggregation and modular chassis aggregation). Although the \ac{xHaul} is not the main focus of this survey, we note that there is a need to integrate the heterogeneity of transport network technologies, into the power modeling for \ac{NG-RAN} for a more accurate representation of the energy impact.

\subsection{Power modeling summary}

This survey covers all the major \ac{RAN} power modeling papers the authors could locate, but it is notable what was missing from the literature. In many ways, it is not unexpected that no papers were found that unambiguously showed the net power consumption for a 5G deployment. Some parameter fitting was done against small deployments of 3G and 4G systems. The most likely explanation is that the information needed to do this is extremely hard to obtain and would be commercially sensitive. In the case of analytical models, it means that necessary parameters are not well-known and, while the power models can be used as part of a modeling package, the uncertainties in the absolute value of the result may be large. In the case of empirical models, it means that unobscured results for 5G systems have, at best, been tested against small test beds. The formal academic literature could not provide even an order of magnitude estimate for the power consumption of a single \ac{BS} in a ``typical" installation. Looking outside formal publications we were able to find two estimates of 1.4kW (for a single band) and 4.7kW (for three or four) for a 5G \ac{BS} under full load, however, it is highly unsatisfactory to resort to such untrustworthy sources. Following such sources further did not seem to fall into the scope of a survey of academic literature. 

\section{Energy Saving Techniques}
\label{sec:energy-saving}
This section investigates the techniques researchers have used to reduce power usage in RAN networks. We categorize techniques into time, spatial, frequency and power domains. We provide a brief overview of approaches within each domain that show promising gains in network energy efficiency. Table~\ref{table:time-domain} provides an at-a-glance summary of the papers considered. Numerical comparison of results between papers was quickly discovered to be insurmountable for a number of reasons. Different authors use different metrics to measure energy saving/efficiency. Some studies allow energy efficiency to be traded against degraded user experience whereas others assume the user experience must remain at least as good. Finally, the studies are done with different modeling assumptions such as path-loss models and layouts of \ac{BS} and \ac{UE}. For these reasons, it is not possible to look at relative gains between two papers and deduce which is better at improving energy efficiency simply by comparing the claimed saving. 

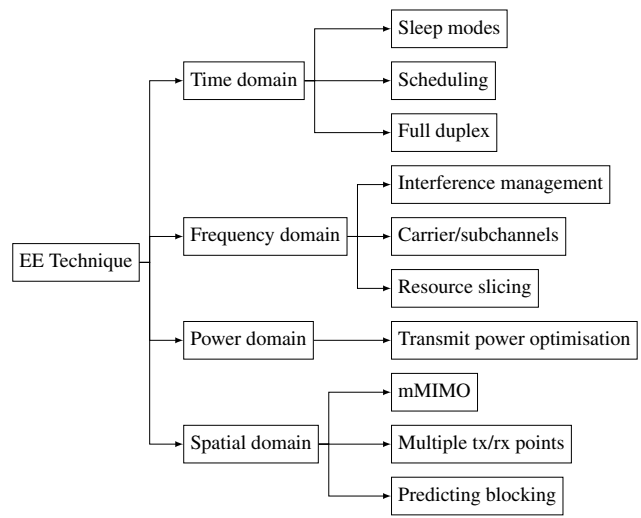
\begin{figure}[ht]
    \centering
    \resizebox{3.3in}{!}{
    \begin{forest}
        for tree={
            grow'=east,        
            forked edges,
            draw, align=center,
            edge={-latex},
            anchor=west,
            child anchor=west,
            l sep+=10pt,        
            tier/.wrap pgfmath arg={tier #1}{level()}
        }
        [EE Technique
            [Time domain
                [Sleep modes]
                [Scheduling]
                [Full duplex]
            ]
            [Frequency domain
                [Interference management]
                [Carrier/subchannels]
                [Resource slicing]
            ]
            [Power domain
                [Transmit power optimisation]
            ]
            [Spatial domain
                [mMIMO]
                [Multiple tx/rx points]
                [Predicting blocking]
            ]
        ]
    \end{forest}
    }
    \caption{Taxonomy of power saving techniques in this survey}
    \label{fig:power_saving}
\end{figure}

Techniques can be split by settings altered in the modeled \ac{RAN} or by the \ac{AI} techniques used to alter those settings. We have split the techniques into four broad areas of resources within wireless communication.
\emph{Time domain} techniques primarily work by moving resources in time. \emph{Frequency domain} techniques optimize by changing the frequencies at which signals are sent. Works based on how \acp{UE}, \ac{BS} and the signals between them are positioned in physical space.
Obviously, some studies will use more than one of these areas in optimization, jointly optimizing transmission power and frequency use. Where a study could fit in more than one section we have tried to fit it according to the main technique used in the primary result presented by the authors. Fig.~\ref{fig:power_saving} shows our taxonomy of techniques based on this split. 

There is a great deal of interest in using the data collected in 5G networks to optimize power reduction and 3GPP~\cite{3gpp_37_817_2022} highlights \ac{AI}/\ac{ML} based solutions to reduce network energy. The techniques they highlight include cell deactivation/sleep (power domain), coverage modification (power and spatial domain) and traffic offloading (spatial domain). New abilities unlocked by 5G enable new techniques that can be used to save energy. For example, having the DU/CU as a virtual appliance removes the dependence on application-specific network hardware, with efficient software implementations that can run on lower-cost general-purpose processing platforms. This also allows functions to dynamically move between hosts and scale computation resources based on the performance requirements or network demand. Scaling with network load helps to improve energy efficiency by dynamically powering down resources during idle periods~\cite{meo_advanced_2021}.

\begin{table*}[ht]
\caption{Summary of EE Techniques}
\label{table:time-domain}
\centering
    \scalebox{0.85}{
    \small
    \begin{tabular}{|c|c|p{3cm}|p{9cm}|} \hline
    {\bf Ref} & {\bf Year} & {\bf Optimization} & {\bf Key Findings or Contributions} \\ \hline
    \multicolumn{4}{|c|}{\bf Time domain techniques}\\ \hline 
    \cite{meo_advanced_2021} 
    & 2021 
    & Stochastic
    & Reduces sleep mode reactivation delay by 90\% for low loads, allowing more time in sleep mode for energy conservation.\\ \hline
    \cite{wu_deep_2021} 
    & 2021 
    & Reinforcement Learning 
    & 20\% reduction in cost of sleep control compared to ARIMA model.\\ \hline
    \cite{amine_energy_2022} 
    & 2022
    & Distributed Q-learning
    & Sleep mode policy optimization with QoS constraints, 5 times more efficient than macro cell offloading. \\ \hline
    \cite{malta_using_2023} 
    & 2023 
    & Reinforcement learning
    & Optimal sleep mode management for trade-off of energy consumption with 5\% energy savings with less than 1 ms latency.\\ \hline
    \cite{iqbal_convolutional_2022} 
    & 2022 
    & Deep Q-network
    & Proposed sleep mode activation saves 5-10\% more power than baseline.\\ \hline
    \cite{li_joint_2021} 
    & 2021 
    & Heuristic
    & Low complexity proactive scheduling algorithm minimizing queuing delay with power constraints.\\ \hline
    \cite{sharara_coordinated_2023} 
    & 2023  
    & Heuristic
    & Reduce wasted power by 48\% from retransmissions by coordinated scheduling of radio and computing resources. \\ \hline
    \cite{hu_profit-based_2021} 
    & 2021 
    & Linear Programming
    & Scheduling algorithms maximize profit margins for network operators.\\ \hline
    \cite{kundu_downlink_2023} 
    & 2023 
    & Analytical
    & Reduce fronthaul capacity demand by duplexing thereby lower energy requirements.\\ \hline
    \cite{nguyen_full-duplex_2022} 
    & 2022 
    & Quadratic Programming
    & Serves multiple users efficiently over long distances using duplexing.\\ \hline
    \multicolumn{4}{|c|}{\bf Frequency domain techniques}\\ \hline 
    \cite{sharma_energy_2023} 
    & 2023 
    & Reinforcement Learning
    & Resource allocation strategy to enhance energy efficiency in ultra-dense networks.\\ \hline
    \cite{yoon_ultima_2023} 
    & 2023
    & Lyapunov framework
    & Optimize interference by sequential user scheduling and power allocation approximation algorithm in Edge SON architecture.\\ \hline
    \cite{marzouk_interference_2022} 
    & 2022
    & Heuristic
    & Resource allocation strategy reduces power consumption up to 93.8\% (low-load) and 64\% (high-load) in \acp{C-RAN}\\ \hline
    \cite{zhou_rate_2021} 
    & 2021
    & Fractional programming
    & Proposed algorithm reduces computing time cost compared to baseline.\\ \hline
    \cite{oliveira_towards_2023} 
    & 2023 
    & Quasi-Newton
    & Resource allocation optimized using second-order optimizer for \ac{RAN} Slicing, compared to dense \ac{NN} approach.\\
    \hline
    \multicolumn{4}{|c|}{\bf Power domain techniques}\\ \hline 
    \cite{kim_deep_2023} 
    & 2023
    & Deep Neural Network
    & Proposed DNN for power allocation reduces computational complexity, while providing similar performance.\\ \hline
    \cite{lee_deep_2023} 
    & 2023 
    & Deep Neural Network
    & Proposed resource allocation scheme achieves 96\% EE of the optimal solution, with reduced computation time.\\ \hline
    \cite{saeidian_downlink_2020} 
    & 2020 
    & Reinforcement Learning
    & Proposed power control algorithm for improved throughput for cell-edge users in urban macro scenarios.\\ \hline
    \cite{alameer_ahmad_power_2021} 
    & 2021 
    & Stochastic
    & Proposed transmit power minimization with imperfect CSIT, serves more users while satisfying minimum QoS constraints.\\ 
    \hline
    \multicolumn{4}{|c|}{\bf Spatial domain techniques}\\ \hline 
    \cite{frenger_massive_2021} 
    & 2021 
    & Analytical
    & Dynamic scaling down of Massive MIMO antenna array size to reduce energy consumption by 30\%\\ \hline
    \cite{marwaha_spatial_2022} 
    & 2022 
    & Analytical
    & Optimizes the number of spatial layers and band activation against user rate requirement while minimizing macro cell power consumption.\\ \hline
    \cite{liu_joint_2021} 
    & 2021  
    & Combinatorial
    & Transmission reception point selection algorithm improves weighted sum energy efficiency in uplink mmWave network over existing methods\\ \hline
    \cite{kim_energy_2022} 
    & 2022 
    & Lyapunov framework
    & Polynomial complexity beam activation and user scheduling algorithm, saves 65\% of average RRH energy consumption in 5G CoMP networks. \\ \hline
    \cite{wu_lidar-aided_2022} 
    & 2022
    & Deep Neural Network
    & LiDAR-aided link blockage prediction enables proactive hand-off with lower delay and more efficient use of network resources \\\hline
    \end{tabular}
    \normalsize
    }
\end{table*}

\subsection{Time-domain}

\textbf{Sleep modes} have naturally attracted a lot of research interest. As mobile networks are designed to handle peak capacity, some base stations remain powered on outside peak hours despite being underutilized. As indicated in \eqref{eq:auer} \acp{SM} are a way to dynamically switch base stations between an active state and power off components in an idle state, depending on user attachment status. This helps to reduce energy consumption in mobile networks by deactivating unnecessary components of the radio transceiver chains when traffic is low. With advanced sleep modes in 5G, a base station is progressively put into deeper sleep modes during increasing periods of inactivity. While deeper sleep modes have lower power consumption, they also introduce longer reactivation delays, impacting user \ac{QoS}. Investigating the compromise between energy saving and delay, the authors in~\cite{meo_advanced_2021} show that dynamic adjustment of the time spent at each SM level can reduce the reactivation delay by 90\% for low loads. They further validate a stochastic model to tune parameters in real time.

Reinforcement learning is a promising approach to achieving real-time optimization. The authors in~\cite{wu_deep_2021} propose a traffic-aware \ac{DRL} based sleep control approach for base stations in large-scale networks using precise mobile traffic forecasting that combines geographical, temporal and semantic spatial (cosine similarity across traffic loads) correlation. They demonstrate that their approach can achieve a 20\% reduction in cost, with energy being the most significant contributing factor, compared to an \ac{ARIMA} forecasting model~\cite{kim2011dynamic}. 

Operating at the \ac{mmWave} spectrum provides wider bandwidth and, therefore, data rates, but signals do not propagate as far, leading to smaller coverage areas. In contrast to macro cells, which cover wider areas, small cells offer increased capacity per geographic area. Small cells are densely deployed to meet the needs of 5G networks. However, overlaps in the coverage area for a small cell in a macro cell region or between neighboring small cells can cause increased interference. In~\cite{amine_energy_2022}, the authors investigate the impact of small cells on the overall performance of 5G networks. They focus on end-users quality of service (QoS) constraints and account for inter-cell interference in a heterogeneous network. The authors propose a distributed Q-learning algorithm that controls the activities of small cells based on their interference level, expected throughput, and buffer size. The findings suggest that, under low traffic loads, moving users from small cells to macro cells can reduce delay and energy consumption for the cluster. However, this may increase the overall network energy consumption at the cost of the average user throughput.

In another reinforcement learning approach~\cite{malta_using_2023}, the authors use a state-action-reward-state-action (SARSA) algorithm to set a sleep mode policy while studying the impact of the wake-up delay of the sleep mode level on the end-to-end user packet latency for uplink traffic. Results show that by increasing the latency threshold to 5~ms and defining low traffic load at 5\%, a 56\% reduction in energy consumption is realized.

In~\cite{iqbal_convolutional_2022}, the authors utilized a \ac{DQN} to conserve 5-10\% more power across all levels of user demand at low loads of up to 50~Mbps. However, the researchers found it unfeasible to meet user quality-of-service demands beyond this point. The \ac{DQN} method involves many state-action pairs, which increases computational complexity and reduces system performance.

\textbf{Scheduling} is crucial in \ac{NG-RAN} to manage increasing data volumes while meeting latency demands and reducing power consumption. Two studies,~\cite{li_joint_2021} and~\cite{sharara_coordinated_2023}, address this challenging task. Researchers in~\cite{li_joint_2021} formulated a Markovian model that efficiently schedules proactive caching and on-demand transmission, analyzing the average delay and power consumption. On the other hand,~\cite{sharara_coordinated_2023} proposed policies that motivate the fact that data processing time strongly depends on the transmission \ac{MCS} index. These policies allow the radio scheduler to set the \ac{MCS} index for users’ transmission based on the radio conditions and the \ac{BBU} pool’s ability to process users’ data. Furthermore, they propose heuristics to reduce power consumption compared to non-coordination heuristics.
Considering the economics of \ac{RAN} operation,
\cite{hu_profit-based_2021} proposed a profit-based algorithm that optimizes task scheduling and resource allocation for \acp{C-RAN} towards maximizing profit margins for network operators. This highlights the need for algorithms adapting to network performance and economic conditions to ensure sustainable growth.

\textbf{Full duplex} systems handle simultaneous data transmission and reception and this has been a catalyst for technological advancements, leading to increased data throughput and enhanced energy efficiency. In~\cite{kundu_downlink_2023}, the authors employed stochastic geometry to analyze a \ac{C-RAN}-enabled full-duplex (FD) cellular network, revealing that strategic \acf{DL} power control significantly boosts the mean rate and mitigates the substantial fronthaul capacity demands in \ac{C-RAN}, thus conserving energy. On the other hand, researchers in~\cite{nguyen_full-duplex_2022} propose a hybrid full-duplex transmission model tailored for 5G networks. The model combines single-mode fiber with free-space optics for \ac{mmWave} signal transmission. It utilizes variable quadrature amplitude modulation to efficiently serve multiple users over long distances.

\subsection{Frequency-domain}

\textbf{Interference Management} (IM) involves avoiding or minimizing interference in a wireless network. In a heterogeneous network (HetNet), small cells (SCs) can generate interference or be affected by interference from a Macro Base Station (MBS) or other nearby SCs. In dense deployments, interference management is crucial for energy efficiency, and intelligent algorithms may help. For example,~\cite{sharma_energy_2023} proposes a reinforcement learning-based resource allocation algorithm to enhance energy efficiency in ultra-dense networks, employing Q-value approximation to tackle the problem of large state spaces and reduce convergence time. Meanwhile, an edge \ac{SON} architecture is proposed in~\cite{yoon_ultima_2023}, integrating centralized and distributed approaches to manage cellular networks with an algorithm that uses Lyapunov optimization for interference management towards performance improvements in real 5G networks. Finally, the work in~\cite{marzouk_interference_2022} focuses on a two-level hybrid resource allocation framework in \acp{C-RAN} that significantly reduces power consumption by up to 93.8\% in low-load conditions, utilizing an admission control algorithm and a heuristic-based RRH-\ac{BBU} mapping algorithm to optimize the number of users and manage interference, considering both \ac{BBU} capacity and user QoS constraints. These studies demonstrate innovative approaches to optimizing network performance while prioritizing energy conservation and effective interference management.

\textbf{Carrier/subchannel} optimization provides another avenue for energy saving.
\Ac{RS} describes the idea where user messages are segmented into common and private components at the transmitter, while at the receiver partial decoding of interference is used and the remainder is treated as noise. With a focus on \ac{RS}, a transmit scheme for \ac{D2D} underlaid cache-enabled \acp{C-RAN} is proposed by the authors in~\cite{zhou_joint_subcarrRS_2022}. They focus on maximizing the sum rate while adhering to power and fronthaul cost constraints through user grouping, dynamic clustering, beamforming, \ac{RS} ratio, and subcarrier allocation. They present algorithms for each subproblem, leading to convergence at a stationary point. The proposed technique achieves a 22\% gain in sum-rate versus \ac{D2D} random scheduling, but the specific amount of energy saved is not quantified.

\textbf{Resource slicing} is a term referring to how resource blocks in a \ac{RAN} are allocated (at heart this involves both the time and frequency domain but here we have included it as a frequency domain technique). 
Conscious of the energy impact of their previous works~\cite{oliveira_mapping_2022}, researchers in~\cite{oliveira_towards_2023} looked at energy optimization using \ac{RAN} slicing. They introduce KPIC-Lite, a neural network-based solution that consumes 700 to 1000 times fewer computational resources than previous models while maintaining performance in most tested scenarios. They offer a new loss function for better convergence and efficient use of a second-order optimizer to reduce computational resource usage. However, the specific energy savings related to \ac{RAN} slicing operations are not explicitly quantified.

\subsection{Power-domain}
\textbf{Transmit power optimization} is a technique to control networks and reduce power consumption. As users move further away from an access point, the transmit power must increase to ensure the signal can reach the receiver with sufficient strength.
However, transmit power reduction strategies must be carefully managed to minimize the loss of \ac{SINR}, which would impact performance within the high density \ac{5G} networks. Recent studies have explored the potential of \ac{AI} to optimize power allocation, such as the deep learning-based resource allocation scheme presented in~\cite{kim_deep_2023}. This scheme includes a subchannel allocation algorithm and a power allocation strategy that uses deep neural networks specifically designed for the \ac{DL} in heterogeneous \ac{NOMA} networks. Another study on ultra-dense small cell networks,~\cite{lee_deep_2023}, proposes a deep learning-based approach to maximize energy efficiency. Their method uses a neural network to determine the activation of small cell base stations, user association, and transmit power. It aims to achieve near-optimal energy management with less than 4 ms computation time across all considered cases, notably within the rigid latency constraints of \ac{3GPP} requirements~\cite{3gpp_servicereqs_2023}.
Another study on dense \ac{5G} networks~\cite{saeidian_downlink_2020} proposes a data-driven approach based on deep reinforcement learning for \ac{DL} power control to improve interference at the cell edge. In contrast to treating interference as an unwanted artifact, the authors in~\cite{alameer_ahmad_power_2021} consider rate splitting for interference mitigation, in addition to their primary focus power of transmit power minimization under imperfect \ac{CSIT}. They show that compared to the conventional ``treating interference as noise'' approach, \ac{RS} uses a lower sum total transmit power for the same number of users in a \ac{C-RAN} system.

\subsection{Spatial-domain}

\textbf{Massive MIMO} provides fine-grained spatial control of signals using multiple antennas. The term `spatial elements' in this context refers to these antennas, a critical part of the system. The \ac{RRC} protocol manages the configuration of these radio resources. This protocol allows for periodic updates to the configurations, enabling changes in the number of antenna ports or elements that are actively used.

\Acf{UE}, such as smartphones, play a vital role by providing the base station with Channel State Information-Reference Signals (CSI-RS). These signals convey the UE's understanding of the channel conditions. Presently, \acp{UE} can support various CSI-RS configurations~\cite{3gpp_38864_2022}, each corresponding to a different quantity of antenna ports or elements. This versatility permits the base station to dynamically adjust which spatial elements are engaged for transmitting data to the UE based on the \ac{CSI} reports, thereby optimizing the communication to suit the prevailing channel conditions. These CSI-RS and corresponding \ac{CSI} reports are tailored to specific segments of the available bandwidth, known as \acp{BWP}.

Towards spatial domain optimization, ~\cite{frenger_massive_2021} propose and evaluate dynamic massive \ac{MIMO} muting, which is a technique that can be used to scale down the active antenna array size when traffic demand is low, hence reducing energy consumption. Whereas the authors in~\cite{marwaha_spatial_2022} propose a spatial and spectral resource allocation for energy-efficient massive \ac{MIMO} \ac{5G} networks. Specifically, they consider spatial optimization by selecting the number of active antennas. The results from~\cite{marwaha_spatial_2022} highlight that a single spatial layer per \ac{PRB} achieves the lowest energy consumption in low-load scenarios.

\textbf{Multiple transmission-reception points} (TRPs) allow optimization of energy efficiency by more intelligently routing radio signals between \ac{UE} and \ac{BS}. Here, there is a capacity to adapt the number of TRPs actively transmitting and receiving signals and channels to a \ac{UE}. Considering this, researchers in~\cite{liu_joint_2021} study the joint transmission reception point (TRP) selection and resource allocation problem to maximize energy efficiency under imperfect channel state information \ac{CSI} for an uplink \ac{mmWave} network. In contrast,~\cite{kim_energy_2022} focuses on the combinatorial beam activation and user scheduling problem; they propose an approximation algorithm to save 65\% of average \ac{RRH} energy consumption for the same average queue backlogs compared to baseline algorithms, which do not consider energy consumption and queue backlog.

\textbf{Predicting signal blockage} can be used to increase efficiency through a better understanding of how signals actually propagate in a physical space. In~\cite{wu_lidar-aided_2022} the authors use LiDAR-aided mobile blockage prediction in real-world \ac{mmWave} systems. Here, spatial elements are considered to predict the physical location and movement of obstacles that can block \ac{LOS} paths. This allows alternate signal paths to be used when a signal path is predicted to be weak. By predicting the blockages with high accuracy, their proactive scheme allows for lower delay and more efficient use of network resources.

\subsection{Summary of AI techniques for RAN efficiency}

We have seen that a large number of levers are available for pulling to increase energy efficiency in a network. We have also seen that a large number of \ac{AI} techniques can be applied to each. A frustration in this survey is the near impossibility of comparing between techniques in the published literature. A lack of reference models and common scenarios makes it irresponsible to compare a claim of x\% saving in one paper with y\% saving in another. While this problem will always remain difficult it could be mitigated by including test scenarios with set parameters that could be replicated between papers as a baseline. However, this relies on those scenarios containing sufficient modeling detail that they can capture the optimization details the researchers wish to model. A further problem is in the reporting of optimization efficiency. The computational requirements of the proposed schemes were orders of magnitude apart but it is unclear how to compare them. Some authors give asymptotic estimates of the algorithmic performance which is a good starting point but certainly not a panacea.

\section{Operational Energy Cost of AI and ML}
\label{sec:ai-ml-energy-ops}
In this section, we review the factors that impact the operational energy cost of \ac{AI} techniques\footnote{As previously discussed, AI here is a catchall and should be considered to include ML.} and consider tools that can help with this. Specifically, we highlight supporting literature for the costs of model inference (as this is the part of the model that will be run continually) considering aspects of data, software and hardware. \ac{AI} forms an essential part of the future of \ac{NG-RAN}~\cite{lin_5g_advanced_2022}, particularly in optimizing network energy usage. The workflow of an \ac{AI} model includes training, testing, and deployment~\cite{amershi_software_2019}. Models deployed in the \ac{NG-RAN} require input of parameters and state information from the local node (e.g. \ac{gNB}), \ac{UE} and neighboring \ac{NG-RAN} nodes~\cite{3gpp_37_817_2022}. The output of the model inference is then used to make predictions or decisions that are hoped to increase the performance and energy efficiency of the RAN. However, an open question is how the energy saving from improved efficiency compares with the energy cost of running the AI/ML pipeline. We will also consider in Section~\ref{sec:ngrancost} the computational costs within the RAN of virtualizing network functions.

\subsection{AI Costs in General}

\begin{table*}[!ht]
\renewcommand{\arraystretch}{1.3}
\caption{AI power consumption factors}
\label{table:ai-ml-power-factors}
\begin{tabular}{|c|m{2.8cm}|m{6cm}|m{6cm}|}
\hline
\textbf{Ref} & \textbf{Factor} & \textbf{Impact} & \textbf{Limitation}\\ \hline

\cite{desislavov_trends_2023}
& Number of parameters
& Correlated, but severity not as strong as anticipated.
& Only computer vision and natural language processing models were evaluated.
\\ \hline

\cite{verdecchia_data-centric_2022}
& Volume of data points
& Approx. 70\% energy saving
& Focuses on data for model training, not inference.
\\ \hline

\cite{verdecchia_systematic_2023}
& Number of inferences
& Multiplicative effect on power consumption.
& Does not account for the power saved by reducing human effort.
\\ \hline

\cite{lin_overview_2023}
& Model updates
& Energy burden from retraining.
& Model portability efficiency unexplored.
\\ \hline

\cite{patterson_carbon_2022}
& Sparsity
& Mixture of experts (MoE) up to 10x lower energy consumption.
& MoE is not widely explored in wireless communication.
\\ \hline
\end{tabular}
\end{table*}

Insights into the costs of running \ac{AI} in the \ac{NG-RAN} context may be gained from studying the costs in a more general setting. A recent study~\cite{wu_sustainable_2022}, reports the energy footprint of data processing of a \acf{RM} accounts for 31\% of the \ac{AI} end-to-end pipeline, based on data center electricity use. In this section, we discuss the salient factors that impact the \ac{AI} power consumption, with a summary provided in Table.~\ref{table:ai-ml-power-factors}, and commentary on the limitations of the studies. 

The computational load of a model is primarily governed by the complexity of the model and the types of operations it must carry out. Models with a larger number of parameters impose a higher computation load to evaluate their relationships, translating into a higher power consumption. 
For example, in~\cite{verdecchia_data-centric_2022}, the authors highlight the impact of modifying datasets to improve energy efficiency of algorithms. Notably, they observe that decreasing the feature set and volume of data points can achieve nearly 70\% energy reduction at a negligible accuracy loss for most algorithms, after factoring in an average of 5\% for data preprocessing overhead. Whereas this study focused on model training, the same group later underscore the gaps in the \ac{AI} pipeline~\cite{verdecchia_systematic_2023}, emphasizing that model training is far less frequent than model inference. One effective way to improve model inference efficiency is an optimizer like \emph{Clover}~\cite{li_clover_2023},
which uses a mix of high and low quality models, alongside GPU partitioning to maintain high accuracy, to match computational load to available resources.
In~\cite{desislavov_trends_2023}, the authors analyze the inference costs of \ac{CV} and \ac{NLP} models, and conclude that energy costs of model inference with respect to the number of parameters are not as rapidly increasing as previously thought~\cite{de_vries_growing_2023}. They attribute this to improvements in both \ac{AI}-optimized hardware and also to efforts that are invested for improving algorithmic efficiency in the years after an algorithm is widely deployed.

Improvements in algorithm design such as model pruning and quantization~\cite{zawishEnergyAI2023} are shown to reduce complexity for fixed energy requirements. Choosing efficient \ac{AI} models, e.g. opting for sparse models, can reduce computation by up to a factor of ten~\cite{patterson_carbon_2022}. Model scalability is improved by training large, sparsely-activated \ac{NN}s~\cite{wu_sustainable_2022}, achieving higher accuracy at a lower operational energy footprint. 
It is worth noting that the model type dictates the start of the inference phase, with supervised models requiring completed training to begin~\cite{3gpp_studyAImano_2023}, unlike reinforcement learning models. In~\cite{lin_overview_2023} the authors note that in the wireless network context, significant changes in system state will often require updates to model parameters and the calculation of new solutions which may have a big impact on energy demand. A consideration for future studies might therefore consider the energy impact of the frequency model inference is executed or the conditions which trigger them that give the greatest benefits.

The potential gains from running computation in a cloud, are explored in~\cite{patterson_carbon_2022}, where experiments show a reduction of computation energy costs by 50\% compared to on-premises. The authors attribute this to the massive investment in the custom design and operation of data centers by cloud providers. However, not all cloud data centers are equally efficient~\cite{mat_brown_digging_2023}. As discussed in Section~\ref{sec:ngrancost}, the choice of datacenters in the \ac{RAN} context will be heavily constrained by latency requirements. 

The execution of an \ac{AI} model with greater computational complexity has a higher energy requirement~\cite{li_clover_2023}, though many other factors (e.g. number of iterations and CPU frequency) may come into play~\cite{pawar_understanding_2020}.
Algorithmic complexity is not always reported and not reported consistently.
Those authors in Section~\ref{sec:energy-saving} who did report algorithmic complexity used Big-O notation, a derivative of \ac{FLOPS} or a custom cost function to do so. This mixture of reporting adds weight to the assertion in~\cite{desislavov_trends_2023} that future studies need detailed and consistent reporting of measures. 

A barrier to reporting may be a lack of awareness of tools available. Tools such as \emph{ptflops}~\cite{ptflops} or \emph{EAIBench}~\cite{zhang_EAIBench_2023} which calculates an energy consumption benchmark for models, do not yet satisfy the need for a robust and mature tool for energy consumption. 
In~\cite{anthony_carbontracker_2020}, the authors highlight a tool that predicts the energy and carbon footprint of DL models that use hardware acceleration, such as \acp{GPU}. They show that the \ac{GPU} consumes approximately 50-60\% of total energy spent during training, with the remaining energy use being the aggregate of CPU and DRAM.
In~\cite{lannelongue_green_2021} the authors present a tool to estimate the carbon footprint of a computational task reliably for a variety of hardware. Their \emph{GreenAlgorithms} calculator calculates the carbon footprint as the product of energy needed and carbon intensity. 

\subsection{Computation Costs for NG-RAN}
\label{sec:ngrancost}
Doing computation in an \ac{NG-RAN} setting has implications that are different from simply doing a similar computation in a general setting. 
The \ac{3GPP} framework has strict latency requirements for 5G systems~\cite{3gpp_servicereqs_2023}. The \ac{3GPP} framework places responsibility for data preparation (e.g. cleaning, formatting and transformation)~\cite{3gpp_37_817_2022}, on the inference model. A related \ac{3GPP} study~\cite{3gpp_studyAImano_2023} on \ac{AI} management emphasizes the importance of selective data usage and the filtering of low-quality data since excessive, irrelevant data increases storage and processing load. 
However, the \ac{O-RAN} Alliance~\cite{oran_wg2_aiml_2021} proposes that \ac{AI} model optimization that is to interact in the Near Real-Time setting must do its computation close (in terms of latency) to the \ac{BS}.
Making computation (\ac{AI} inference and/or data processing) either co-located with the BS or physically distributed has drawbacks such as loss of the efficiencies of cloud computing in a datacenter (see Section~\ref{sec:ai-ml-energy-ops}).

Studies~\cite{rost_complexityrate_2015, nikaein_processing_2015}, estimate the computation requirements of BBU functions with respect to parameters such as \ac{MCS}, \ac{SNR} and bandwidth.
The authors in~\cite{ge_energy_2017} argue that \ac{BBU} processing will significantly impact energy consumption when considering densely deployed small cells with low transmit power in a 5G network as a replacement for high-power macro cells with larger bandwidths. The study uses Landauer's principle~\cite{landauer_irreversibility_1961} to model the computational power of a small cell and macro cell \ac{BBU}. Simulations reveal the computational power ratio (computation power required over the total power at a base station) of 5G small cells is more than 50\%. However, more recently~\cite{wesemann_energy_2023} highlights the trend for integration of \acf{BB} functions into the \ac{RU}, wherein latency critical parts of \ac{BB} processing reside in at the radio site and delay tolerant processing may be offloaded to virtualized \acp{DU} or \acp{CU}. 

As previously mentioned, the move to \acp{vBS} means we need to understand the power consumed by modeling those functions that are virtualized. A recent study~\cite{khatibi_modelling_2018} models the computational power needed to provision a virtual RAN. The authors note that the share of \ac{BBU} processing time is limited to 3~ms per subframe, owing to the standardized \ac{HARQ} feature. As a result, they argue that virtual resources must be sufficiently allocated to meet this requirement and provide a model to approximate the processing time based on the \ac{CPU} frequency, number of \ac{PRB} and \ac{MCS} index. Based on an \ac{O-RAN} testbed, they show that a \ac{CPU} with advanced vector extension support, requires a minimum clock speed of 2 GHz. Looking at the virtualized \ac{BS} as a whole, the authors in~\cite{ayala-romero_experimental_2021} study how \ac{SNR}, \ac{MCS} and bandwidth parameters affect \ac{CPU} power consumption of a \ac{GPP}. Observing nonlinear effects of \ac{SNR} on power consumption, the authors explain that higher noise necessitates more iterations of the turbo decoder to process the signal thereby increasing the computational load.

Lightweight models tailored specifically to \ac{NG-RAN} needs could reduce the energy consumed by \ac{AI} models. For example, in~\cite{palitharathnaSLIVER2023}, the authors use a \emph{block-wise} training approach to reduce the complexity of the path and orientation prediction, with subsequent \ac{IRS} angle optimization and beamforming. Simulations show the proposed \ac{ANN} can reduce transmit power by up to 40\% with two \acp{IRS} in a system with 10 energy users. Moreover, these results serve as an example of how wise choices in \ac{ML} design can help to ease the impact on \ac{NG-RAN} power consumption through efficient computation methods. 

\subsection{Future Research Opportunities}
\label{sec:future-research}
The research on energy usage of \ac{AI}, particularly within the \ac{RAN} context, highlighted some significant research gaps. It is not using the current literature to compare the energy cost of running \ac{AI} models for energy efficiency compared with the energy savings created by those models. Studies that use \ac{AI} for energy reduction do not consistently report complexity and do not take advantage of existing tools that can estimate energy usage. Most give little attention to the trade-off between model complexity (and hence energy consumption) and inference accuracy. One possible solution relies on the report card approach outlined in~\cite{castanoHugFaceCarbon2023}. This approach has been used to report the carbon emissions with \ac{NLP} applications leading the reporting. 

Most research encountered did not align with industry standards and specifications particularly those outlined by standard bodies like the \ac{3GPP} and \ac{O-RAN}. For example, solutions need to meet certain latency requirements and it was far from clear that this was generally the case. While there was a body of research into the efficiency gains of \ac{vBS} and optimizations that this makes possible, there is a lack of clarity on the energy requirements from that virtualization. 

\section{Conclusions}
\label{sec:conclusions}

This paper examined the literature related to energy efficiency in next-generation mobile networks, focusing on the \ac{RAN}. To this end, the power consumption of the RAN is first studied to form a baseline understanding of the power consumption of different functions in the RAN and how these vary with different traffic loads. Energy efficiency in the RAN is often addressed by optimizing RAN resources; this paper proposed four categorization to analyses this process. These are defined based on the location of the degrees of freedom in the optimization process including time, frequency, power, and space. Recent advances and successes of \ac{AI} have led to a surge in research that employs \ac{AI} to address these optimization objectives which are often impossible to solve analytically. 

The first challenge encountered in this study is the difficulty of conducting a fair and correct comparison of power saving capabilities of the surveyed works. This is due to the lack of reference models and common test scenarios with set parameters that would be used to rank the power-saving capabilities of each of these works. 

Another gap found in the literature is the lack or inconsistent reporting of \ac{AI} complexity, and henceforth power consumption, thus the failure to answer the question of ``At what \ac{AI} power consumption cost is RAN power saving of the proposed method achieved and is worthwhile?''

Despite a recent increase in awareness of the power consumption of general \ac{AI}, a great research gap remains in its application to RAN energy efficiency optimization and, therefore, in the energy-aware design of \ac{AI} for RAN. The degrees of freedom in the RAN energy-efficiency problem are limited by rigid requirements from the standards such as latency and quality of service which adds constraints to the design of energy-aware \ac{AI}.

There is no denying that \ac{AI} has a critical role to play in the evolution of mobile networks as recognized by standardization groups (e.g., \ac{3GPP} and \ac{O-RAN}) nonetheless, a significant amount of research is needed to bridge the gaps identified in this paper to drive an effective and efficient pathway for this role.

\section*{Acknowledgment}
The authors thank Keith Briggs for contributions to editing.

\bibliographystyle{IEEEtran}
\bibliography{IEEEabrv,refs}

\begin{thebibliography}{100}
\providecommand{\url}[1]{#1}
\csname url@samestyle\endcsname
\providecommand{\newblock}{\relax}
\providecommand{\bibinfo}[2]{#2}
\providecommand{\BIBentrySTDinterwordspacing}{\spaceskip=0pt\relax}
\providecommand{\BIBentryALTinterwordstretchfactor}{4}
\providecommand{\BIBentryALTinterwordspacing}{\spaceskip=\fontdimen2\font plus
\BIBentryALTinterwordstretchfactor\fontdimen3\font minus \fontdimen4\font\relax}
\providecommand{\BIBforeignlanguage}[2]{{%
\expandafter\ifx\csname l@#1\endcsname\relax
\typeout{** WARNING: IEEEtran.bst: No hyphenation pattern has been}%
\typeout{** loaded for the language `#1'. Using the pattern for}%
\typeout{** the default language instead.}%
\else
\language=\csname l@#1\endcsname
\fi
#2}}
\providecommand{\BIBdecl}{\relax}
\BIBdecl

\bibitem{gsma_mobile_2023}
{GSMA}, ``Mobile net zero: {S}tate of the industry on climate action 2023,'' Feb. 2023.

\bibitem{soldati_approaching_2023}
P.~Soldati, E.~Ghadimi, B.~Demirel, Y.~Wang, M.~Sintorn, and R.~Gaigalas, ``Approaching {AI}-native {RAN}s through generalization and scalability of learning,'' \emph{Ericsson Technology Review}, vol. 2023, no.~3, pp. 2--12, Mar. 2023.

\bibitem{de_vries_growing_2023}
A.~de~Vries, ``The growing energy footprint of artificial intelligence,'' \emph{Joule}, vol.~7, no.~10, pp. 2191--2194, Oct. 2023.

\bibitem{wu_sustainable_2022}
C.-J. Wu \emph{et~al.}, ``Sustainable {AI}: Environmental implications, challenges and opportunities,'' in \emph{Proc. of the 5th MLSys Conf.}, Santa Clara, CA, USA, Jan. 2022.

\bibitem{patterson_carbon_2022}
D.~Patterson \emph{et~al.}, ``The carbon footprint of machine learning training will plateau, then shrink,'' \emph{Computer}, vol.~55, no.~7, pp. 18--28, Jul. 2022.

\bibitem{rodoshi_resource_2020}
R.~T. Rodoshi, T.~Kim, and W.~Choi, ``Resource management in cloud radio access network: Conventional and new approaches,'' \emph{Sensors}, vol.~20, no.~9, p. 2708, May 2020.

\bibitem{lopez-perez_survey_2022}
D.~L\'{o}pez-P\'{e}rez \emph{et~al.}, ``A survey on {5G} radio access network energy efficiency: Massive {MIMO}, lean carrier design, sleep modes, and machine learning,'' \emph{{IEEE} Commun. Surv. Tuts.}, vol.~24, no.~1, pp. 653--697, Jan. 2022.

\bibitem{brik_deep_2022}
B.~Brik, K.~Boutiba, and A.~Ksentini, ``Deep learning for {B5G} {O}pen radio access network: Evolution, survey, case studies, and challenges,'' \emph{{IEEE} Open J. Commun. Soc}, vol.~3, pp. 228--250, Jan. 2022.

\bibitem{abubakar_energy_2023}
A.~I. Abubakar, O.~Onireti, Y.~Sambo, L.~Zhang, G.~K. Ragesh, and M.~Ali~Imran, ``Energy efficiency of {O}pen radio access network: A survey,'' in \emph{{IEEE} 97th {Veh} {Technol.} {Conf.}}, Florence, Italy, Jun. 2023, pp. 1--7.

\bibitem{larsen_toward_2023}
L.~M. Larsen, H.~L. Christiansen, S.~Ruepp, and M.~S. Berger, ``Toward greener {5G} and beyond radio access networks--{A} survey,'' \emph{{IEEE} Open J. Commun. Soc}, vol.~4, pp. 768--797, Mar. 2023.

\bibitem{mao_ai_2022}
B.~Mao, F.~Tang, Y.~Kawamoto, and N.~Kato, ``{AI} models for {Green Communications} towards {6G},'' \emph{{IEEE} Commun. Surv. Tuts.}, vol.~24, no.~1, pp. 210--247, Feb. 2022.

\bibitem{garcia-martin_estimation_2019}
E.~Garc\'{\i}a-Mart\'{\i}n, C.~F. Rodrigues, G.~Riley, and H.~Grahn, ``Estimation of energy consumption in machine learning,'' \emph{Journal of Parallel and Distributed Computing}, vol. 134, pp. 75--88, Dec. 2019.

\bibitem{azariEvolutionNTN2022}
M.~M. Azari \emph{et~al.}, ``Evolution of non-terrestrial networks from {5G} to {6G}: A survey,'' \emph{{IEEE} Communications Surveys \& Tutorials}, vol.~24, no.~4, pp. 2633--2672, Nov. 2022.

\bibitem{palitharathnaSLIVER2023}
K.~W.~S. Palitharathna, A.~Maria~Vegni, and H.~A. Suraweera, ``{SLIVER}: A {SLIPT}-enabled {IRS}-assisted {VLC} system for energy optimization,'' in \emph{2023 {IEEE} 20th International Conference on Mobile Ad Hoc and Smart Systems ({MASS})}.\hskip 1em plus 0.5em minus 0.4em\relax {IEEE}, Sep. 2023, pp. 143--151.

\bibitem{papanikolaouSimultaneous2024}
V.~K. Papanikolaou \emph{et~al.}, ``Simultaneous lightwave information and power transfer in {6G} networks,'' \emph{{IEEE} Communications Magazine}, vol.~62, no.~3, pp. 16--22, Mar. 2024.

\bibitem{wu_lidar-aided_2022}
S.~Wu, C.~Chakrabarti, and A.~Alkhateeb, ``{LiDAR}-aided mobile blockage prediction in real-world millimeter {Wave} systems,'' in \emph{{IEEE} {Wireless} {Commun.} and {Netw.} {Conf.}}, Austin, TX, USA, Apr. 2022, pp. 2631--2636.

\bibitem{rahimMultiIRSAidedTerahertzNetworks2024}
M.~Rahim, T.~L. Nguyen, G.~Kaddoum, and T.~N. Do, ``Multi-{IRS}-aided terahertz networks: Channel modeling and user association with imperfect {CSI},'' \emph{{IEEE} Open Journal of the Communications Society}, vol.~5, pp. 836--855, Jan. 2024.

\bibitem{3gpp_NG-RAN_Overall_Desc_ts38_300_v17_6_0}
{3GPP}, ``{NR} and {NG-RAN} overall description ({Release} 17),'' Oct. 2023.

\bibitem{3gpp_arch_desc_ts_38.401_v17.6.0}
------, ``Architecture description ({Release} 17),'' Oct. 2023.

\bibitem{arzo_study_2020}
S.~T. Arzo, R.~Bassoli, F.~Granelli, and F.~H.~P. Fitzek, ``Study of virtual network function placement in {5G} cloud radio access network,'' \emph{{IEEE} Trans. Netw. Service Manag.}, vol.~17, no.~4, pp. 2242--2259, Dec. 2020.

\bibitem{gkatzios_optimized_2020}
N.~Gkatzios, M.~Anastasopoulos, A.~Tzanakaki, and D.~Simeonidou, ``Optimized placement of virtualized resources for {5G} services exploiting live migration,'' \emph{Photon. Netw. Commun.}, vol.~40, no.~3, pp. 233--244, Dec. 2020.

\bibitem{moreira_zorello_power-efficient_2022}
L.~M. Moreira~Zorello, M.~Sodano, S.~Troia, and G.~Maier, ``Power-efficient baseband-function placement in latency-constrained {5G} metro access,'' \emph{{IEEE} Trans. Green Commun. Netw.}, vol.~6, no.~3, pp. 1683--1696, Sep. 2022.

\bibitem{amiri_energy-aware_2023}
E.~Amiri, N.~Wang, M.~Shojafar, and R.~Tafazolli, ``Energy-aware dynamic {VNF} splitting in {O-RAN} using deep reinforcement learning,'' \emph{IEEE Wireless Commun. Lett.}, vol.~12, no.~11, pp. 1891--1895, Nov. 2023.

\bibitem{wesemann_energy_2023}
S.~Wesemann, J.~Du, and H.~Viswanathan, ``Energy efficient extreme {MIMO}: Design goals and directions,'' \emph{{IEEE} Commun. Mag.}, vol.~61, no.~10, pp. 132--138, Oct. 2023.

\bibitem{auer_how_2011}
G.~Auer \emph{et~al.}, ``How much energy is needed to run a wireless network?'' \emph{{IEEE} Wireless Commun.}, vol.~18, no.~5, pp. 40--49, Oct. 2011.

\bibitem{bjornson_massive_2017}
E.~Bj\"{o}rnson, J.~Hoydis, and L.~Sanguinetti, ``Massive {MIMO} networks: {Spectral}, energy, and hardware efficiency,'' \emph{Foundations and Trends in Signal Processing}, vol.~11, no. 3-4, pp. 154--655, Nov. 2017.

\bibitem{holtkamp_parameterized_2013}
H.~Holtkamp, G.~Auer, V.~Giannini, and H.~Haas, ``A parameterized base station power model,'' \emph{{IEEE} Commun. Lett.}, vol.~17, no.~11, pp. 2033--2035, Nov. 2013.

\bibitem{desset_towards_2013}
C.~Desset, B.~Debaillie, and F.~Louagie, ``Towards a flexible and future-proof power model for cellular base stations,'' in \emph{24th {T}yrrhenian {Int.} {Workshop} on {Digit.} {Commun.}}, Genoa, Italy, Sep. 2013, pp. 1--6.

\bibitem{b_debaillie_flexible_2015}
{B. Debaillie}, {C. Desset}, and {F. Louagie}, ``A flexible and future-proof power model for cellular base stations,'' in \emph{{IEEE} 81st {Veh} {Technol.} {Conf.}}, May 2015, pp. 1--7.

\bibitem{bjornson_optimal_2015}
E.~Bjornson, L.~Sanguinetti, J.~Hoydis, and M.~Debbah, ``Optimal design of energy-efficient multi-user {MIMO} systems: Is massive {MIMO} the answer?'' \emph{{IEEE} Trans. Wireless Commun.}, vol.~14, no.~6, pp. 3059--3075, Jun. 2015.

\bibitem{eramo_trade-off_2016}
V.~Eramo, M.~Listanti, F.~G. Lavacca, P.~Iovanna, G.~Bottari, and F.~Ponzini, ``Trade-off between power and bandwidth consumption in a reconfigurable {xHaul} network architecture,'' \emph{IEEE Access}, vol.~4, pp. 9053--9065, Dec. 2016.

\bibitem{larsen_ran_2022}
L.~M. Larsen, S.~Ruepp, M.~S. Berger, and H.~L. Christiansen, ``{RAN} design guidelines for energy efficient {5G} mobile {xHaul} networks,'' in \emph{14th {Int.} {Conf.} on {Commun.} ({COMM})}, Bucharest, Romania, Jun. 2022, pp. 1--6.

\bibitem{capone_modeling_2017}
A.~Capone, S.~D'Elia, I.~Filippini, A.~E.~C. Redondi, and M.~Zangani, ``Modeling energy consumption of mobile radio networks: An operator perspective,'' \emph{{IEEE} Wireless Commun.}, vol.~24, no.~4, pp. 120--126, Aug. 2017.

\bibitem{piovesan_machine_2022}
N.~Piovesan, D.~L\'{o}pez-P\'{e}rez, A.~De~Domenico, X.~Geng, H.~Bao, and M.~Debbah, ``Machine learning and analytical power consumption models for {5G} base stations,'' \emph{{IEEE} Commun. Mag.}, vol.~60, no.~10, pp. 56--62, Oct. 2022.

\bibitem{huttunen_base_2023}
J.~Huttunen, M.~Parssinen, T.~Heikkila, O.~Salmela, J.~Manner, and E.~Pongracz, ``Base station energy use in dense urban and suburban areas,'' \emph{IEEE Access}, vol.~11, pp. 2863--2874, Jan. 2023.

\bibitem{dappworks_front_2020}
\BIBentryALTinterwordspacing
Dappworks. (2020, May) Front line data study about {5G} power consumption. [Online]. Available: \url{https://dappworks.com/front-line-data-study-about-5g-power-consumption-you-need-to-know-about-5g}
\BIBentrySTDinterwordspacing

\bibitem{chen_dongxu_5g_2020}
\BIBentryALTinterwordspacing
C.~Dongxu. (2020, Jul.) {5G} power: Creating a green grid that slashes costs, emissions \& energy use. [Online]. Available: \url{https://www.huawei.com/en/huaweitech/publication/89/5g-power-green-grid-slashes-costs-emissions-energy-use}
\BIBentrySTDinterwordspacing

\bibitem{gartner_gartner_2023}
\BIBentryALTinterwordspacing
Gartner, ``Gartner says worldwide {PC} shipments declined 28.5\% in fourth quarter of 2022 and 16.2\% for the year,'' Jan. 2023. [Online]. Available: \url{https://www.gartner.com/en/newsroom/press-releases/2023-01-11-gartner-says-worldwide-pc-shipments-declined-28-\percent-in-fourth-quarter-of-2022-and-16-percent-for-the-year}
\BIBentrySTDinterwordspacing

\bibitem{lenovo_thinkstation_2023}
Lenovo, ``{ThinkStation P3 Ultra} user guide,'' Sep. 2023.

\bibitem{hpe_hpe_2024}
{HPE}, ``{HPE ProLiant DL110 Gen10 Plus Telco} server data sheet,'' Jan. 2024.

\bibitem{supermicro_5g_2024}
Supermicro, ``{5G DU SYS-111E-FWTR-1U} specsheet,'' Jan. 2024.

\bibitem{rodriguez_dell_2023}
E.~Rodriguez, ``{DELL PowerEdge XR8000r} product environmental compliance,'' {DELL T}echnologies, Datasheet, May 2023.

\bibitem{ayala-romero_experimental_2021}
J.~A. Ayala-Romero, I.~Khalid, A.~Garcia-Saavedra, X.~Costa-Perez, and G.~Iosifidis, ``Experimental evaluation of power consumption in virtualized base stations,'' in \emph{{IEEE} {Int.} {Conf.} on {Commun.}}, Montreal, QC, Canada, Jun. 2021, pp. 1--6.

\bibitem{salvat_open_2023}
J.~X. Salvat, J.~A. Ayala-Romero, L.~Zanzi, A.~Garcia-Saavedra, and X.~Costa-Perez, ``Open radio access networks ({O-RAN}) experimentation platform: Design and datasets,'' \emph{{IEEE} Commun. Mag.}, vol.~61, no.~9, pp. 138--144, Sep. 2023.

\bibitem{katsaros_power_2022}
G.~N. Katsaros, R.~Tafazolli, and K.~Nikitopoulos, ``On the power consumption of massive-{MIMO}, {5G} {New Radio} with software-based {PHY} processing,'' in \emph{{IEEE} {Globecom} {Workshops}}, Rio de Janeiro, Brazil, Dec. 2022, pp. 765--770.

\bibitem{matoussi_5g_2020}
S.~Matoussi, I.~Fajjari, S.~Costanzo, N.~Aitsaadi, and R.~Langar, ``{5G} {RAN}: Functional split orchestration optimization,'' \emph{IEEE Journal on Selected Areas in Communications}, vol.~38, no.~7, pp. 1448--1463, Jul. 2020.

\bibitem{pawar_understanding_2020}
U.~Pawar, A.~K. Singh, K.~Malde, B.~R. Tamma, and A.~Antony~Franklin, ``Understanding energy consumption of cloud radio access networks: an experimental study,'' in \emph{{IEEE} 3rd {5G} {World} {Forum} ({5GWF})}, Bangalore, India, Sep. 2020, pp. 407--412.

\bibitem{baliga_energy_2011}
J.~Baliga, R.~Ayre, K.~Hinton, and R.~Tucker, ``Energy consumption in wired and wireless access networks,'' \emph{{IEEE} Commun. Mag.}, vol.~49, no.~6, pp. 70--77, Jun. 2011.

\bibitem{granell_energy_2012}
E.~Granell, S.~Andrade-Morelli, E.~Ruiz-Sanchez, and J.~Lloret, ``Energy consumption study of network access switches to enhance energy distribution,'' in \emph{{IEEE} {Globecom} {Workshops}}, Anaheim, CA, USA, Dec. 2012, pp. 1496--1501.

\bibitem{vishwanath_member_modeling_2014}
A.~Vishwanath, K.~Hinton, R.~W.~A. Ayre, and R.~S. Tucker, ``Modeling energy consumption in high-capacity routers and switches,'' \emph{IEEE Journal on Selected Areas in Communications}, vol.~32, no.~8, pp. 1524--1532, Aug. 2014.

\bibitem{francini_low-cost_2015}
A.~Francini, S.~Fortune, T.~Klein, and M.~Ricca, ``A low-cost methodology for profiling the power consumption of network equipment,'' \emph{{IEEE} Commun. Mag.}, vol.~53, no.~5, pp. 250--256, May 2015.

\bibitem{3gpp_37_817_2022}
{3GPP}, ``Study on enhancement for data collection for {NR} and {EN-DC},'' Apr. 2022, {TR} 37.817.

\bibitem{meo_advanced_2021}
M.~Meo, D.~Renga, and Z.~Umar, ``Advanced sleep modes to comply with delay constraints in energy efficient {5G} networks,'' in \emph{{IEEE} 93rd {Veh.} {Technol.} {Conf.}}, Helsinki, Finland, Apr. 2021, pp. 1--7.

\bibitem{wu_deep_2021}
Q.~Wu, X.~Chen, Z.~Zhou, L.~Chen, and J.~Zhang, ``Deep reinforcement learning with spatio-temporal traffic forecasting for data-driven base station sleep control,'' \emph{IEEE/ACM Transactions on Networking}, vol.~29, no.~2, pp. 935--948, Apr. 2021.

\bibitem{amine_energy_2022}
A.~E. Amine, J.-P. Chaiban, H.~A.~H. Hassan, P.~Dini, L.~Nuaymi, and R.~Achkar, ``Energy optimization with multi-sleeping control in {5G} heterogeneous networks using reinforcement learning,'' \emph{{IEEE} Trans. Netw. Service Manag.}, vol.~19, no.~4, pp. 4310--4322, Dec. 2022.

\bibitem{malta_using_2023}
S.~Malta, P.~Pinto, and M.~Fernandez-Veiga, ``Using reinforcement learning to reduce energy consumption of ultra-dense networks with {5G} use cases requirements,'' \emph{IEEE Access}, vol.~11, pp. 5417--5428, Jan. 2023.

\bibitem{iqbal_convolutional_2022}
A.~Iqbal, M.-L. Tham, and Y.~C. Chang, ``Convolutional neural network-based deep {Q}-network {(CNN-DQN)} resource management in cloud radio access network,'' \emph{China Communications}, vol.~19, no.~10, pp. 129--142, Oct. 2022.

\bibitem{li_joint_2021}
C.~Li, W.~Chen, and K.~B. Letaief, ``Joint scheduling of proactive caching and on-demand transmission traffics over shared spectrum,'' \emph{IEEE Transactions on Communications}, vol.~69, no.~12, pp. 8319--8334, Dec. 2021.

\bibitem{sharara_coordinated_2023}
M.~Sharara, S.~Hoteit, P.~Brown, and V.~V\`{e}que, ``On coordinated scheduling of radio and computing resources in cloud-{RAN},'' \emph{{IEEE} Trans. Netw. Service Manag.}, vol.~20, no.~3, pp. 2990--3003, Sep. 2023.

\bibitem{hu_profit-based_2021}
C.-C. Hu, ``Profit-based algorithm of joint real-time task scheduling and resource allocation in {C-RAN}s,'' \emph{IEEE Internet of Things Journal}, vol.~8, no.~2, pp. 941--950, Jan. 2021.

\bibitem{kundu_downlink_2023}
A.~M. Kundu and T.~V. Sreejith, ``Downlink power control in {C-RAN} enabled full duplex cellular networks,'' \emph{Physical Communication}, vol.~60, p. 102154, Oct. 2023.

\bibitem{nguyen_full-duplex_2022}
D.-N. Nguyen \emph{et~al.}, ``Full-duplex transmission of multi-{Gb/s} subcarrier multiplexing and {5G} {NR} signals in 39 {GHz} band over fiber and space,'' \emph{Applied Optics}, vol.~61, no.~5, p. 1183, Feb. 2022.

\bibitem{sharma_energy_2023}
N.~Sharma and K.~Kumar, ``Energy efficient clustering and resource allocation strategy for ultra-dense networks: A machine learning framework,'' \emph{{IEEE} Trans. Netw. Service Manag.}, vol.~20, no.~2, pp. 1884--1897, Jun. 2023.

\bibitem{yoon_ultima_2023}
P.~Yoon, J.~Hong, S.~Ahn, Y.~Cho, J.~Na, and J.~Kwak, ``{ULTIMA}: Ultimate balance of centralized and distributed benefits for interference management in {5G} cellular networks,'' \emph{IEEE Access}, vol.~11, pp. 85\,694--85\,710, Aug. 2023.

\bibitem{marzouk_interference_2022}
F.~Marzouk, J.~P. Barraca, and A.~Radwan, ``Interference and {QoS}-aware resource allocation considering {DAS} behavior for {C-RAN} power minimization,'' \emph{IEEE Canadian Journal of Electrical and {Comput.} Engineering}, vol.~45, no.~4, pp. 442--453, Dec. 2022.

\bibitem{zhou_rate_2021}
J.~Zhou, Y.~Sun, R.~Chen, and C.~Tellambura, ``Rate splitting multiple access for multigroup multicast beamforming in cache-enabled {C-RAN},'' \emph{IEEE Transactions on Vehicular Technology}, vol.~70, no.~12, pp. 12\,758--12\,770, Dec. 2021.

\bibitem{oliveira_towards_2023}
A.~Oliveira and T.~Vaz\~{a}o, ``Towards green machine learning for resource allocation in beyond {5G} {RAN} slicing,'' \emph{{Comput.} Networks}, vol. 233, p. 109877, Sep. 2023.

\bibitem{kim_deep_2023}
D.~Kim, S.~Kwon, H.~Jung, and I.-H. Lee, ``Deep learning-based resource allocation scheme for heterogeneous {NOMA} networks,'' \emph{IEEE Access}, vol.~11, pp. 89\,423--89\,432, Aug. 2023.

\bibitem{lee_deep_2023}
W.~Lee, H.~Lee, and H.-H. Choi, ``Deep learning-based network-wide energy efficiency optimization in ultra-dense small cell networks,'' \emph{IEEE Transactions on Vehicular Technology}, vol.~72, no.~6, pp. 8244--8249, Jun. 2023.

\bibitem{saeidian_downlink_2020}
S.~Saeidian, S.~Tayamon, and E.~Ghadimi, ``Downlink power control in dense {5G} radio access networks through deep reinforcement learning,'' in \emph{{IEEE} {Int.} {Conf.} on {Commun.}}, Dublin, Ireland, Jun. 2020, pp. 1--6.

\bibitem{alameer_ahmad_power_2021}
A.~Alameer~Ahmad \emph{et~al.}, ``Power minimization using rate splitting with statistical {CSI} in cloud-radio access networks,'' \emph{Frontiers in Commun. and Netw.}, vol.~2, pp. 1--19, Sep. 2021.

\bibitem{frenger_massive_2021}
P.~Frenger and K.~W. Helmersson, ``Massive {MIMO} muting using dual-polarized and array-size invariant beamforming,'' in \emph{{IEEE} 93rd {Veh} {Technol.} {Conf.} ({VTC2021}-{Spring})}, Helsinki, Finland, Apr. 2021, pp. 1--6.

\bibitem{marwaha_spatial_2022}
S.~Marwaha, E.~A. Jorswieck, D.~Lopez-Perez, X.~Geng, and H.~Bao, ``Spatial and spectral resource allocation for energy-efficient massive {MIMO} {5G} networks,'' in \emph{{IEEE} {Int.} {Conf.} on {Commun.}}, Seoul, Korea, Republic of, May 2022, pp. 135--140.

\bibitem{liu_joint_2021}
Y.~Liu, X.~Fang, and M.~Xiao, ``Joint transmission reception point selection and resource allocation for energy-efficient millimeter-{Wave} communications,'' \emph{IEEE Transactions on Vehicular Technology}, vol.~70, no.~1, pp. 412--428, Jan. 2021.

\bibitem{kim_energy_2022}
Y.~Kim, J.~Jeong, S.~Ahn, J.~Kwak, and S.~Chong, ``Energy and delay guaranteed joint beam and user scheduling policy in {5G} {CoMP} networks,'' \emph{{IEEE} Trans. Wireless Commun.}, vol.~21, no.~4, pp. 2742--2756, Apr. 2022.

\bibitem{kim2011dynamic}
H.-W. Kim, J.-H. Lee, Y.-H. Choi, Y.-U. Chung, and H.~Lee, ``Dynamic bandwidth provisioning using {ARIMA}-based traffic forecasting for mobile {WiMAX},'' \emph{{Comput.} Communications}, vol.~34, no.~1, pp. 99--106, Jan. 2011.

\bibitem{zhou_joint_subcarrRS_2022}
J.~Zhou, Y.~Sun, C.~Tellambura, and G.~Y. Li, ``Joint user grouping, sparse beamforming, and subcarrier allocation for {D2D} underlaid cache-enabled {C-RAN}s with rate splitting,'' \emph{IEEE Transactions on Vehicular Technology}, vol.~71, no.~4, pp. 3792--3806, Apr. 2022.

\bibitem{oliveira_mapping_2022}
A.~Oliveira and T.~Vazao, ``Mapping network performance to radio resources,'' in \emph{{Int.} {Conf.} on {Inf.} {Netw.} ({ICOIN})}.\hskip 1em plus 0.5em minus 0.4em\relax Los Alamitos, CA, USA: IEEE Computer Society, Jan. 2022, pp. 298--303.

\bibitem{3gpp_servicereqs_2023}
{3GPP}, ``Service requirements for the {5G} system ({Release} 18),'' 3GPP, TS 22.261 v18.12.0, Dec. 2023.

\bibitem{3gpp_38864_2022}
------, ``Study on network energy savings for {NR},'' 3GPP, TR 38.864, Dec. 2022.

\bibitem{lin_5g_advanced_2022}
X.~Lin, ``An overview of {5G} {A}dvanced evolution in {3GPP} {Release} 18,'' \emph{IEEE Communications Standards Magazine}, vol.~6, no.~3, pp. 77--83, Sep. 2022.

\bibitem{amershi_software_2019}
S.~Amershi \emph{et~al.}, ``Software engineering for machine learning: A case study,'' in \emph{{IEEE}/{ACM} 41st {Int.} {Conf.} on {Softw.} {Eng.}}, Montreal, QC, Canada, May 2019, pp. 291--300.

\bibitem{desislavov_trends_2023}
R.~Desislavov, F.~Mart\'{\i}nez-Plumed, and J.~Hern\'{a}ndez-Orallo, ``Trends in {AI} inference energy consumption: Beyond the performance-vs-parameter laws of deep learning,'' \emph{Sustain. Comput.: Inform. and Syst.}, vol.~38, p. 100857, Apr. 2023.

\bibitem{verdecchia_data-centric_2022}
R.~Verdecchia, L.~Cruz, J.~Sallou, M.~Lin, J.~Wickenden, and E.~Hotellier, ``Data-centric {Green} {AI} an exploratory empirical study,'' in \emph{{Int.} {Conf.} on {ICT} for {Sustain.}}, Plovdiv, Bulgaria, Jun. 2022, pp. 35--45.

\bibitem{verdecchia_systematic_2023}
R.~Verdecchia, J.~Sallou, and L.~Cruz, ``A systematic review of {Green} {AI},'' \emph{WIREs Data Mining and Knowl. Discovery}, vol.~13, no.~4, p. e1507, Jul. 2023.

\bibitem{lin_overview_2023}
X.~Lin, ``An overview of the {3GPP} study on artificial intelligence for {5G} {New Radio},'' Aug. 2023, arXiv preprint arXiv:2308.05315.

\bibitem{li_clover_2023}
B.~Li, S.~Samsi, V.~Gadepally, and D.~Tiwari, ``Clover: Toward sustainable {AI} with carbon-aware machine learning inference service,'' in \emph{Proc. of the {Int.} {Conf.} for {High} {Perform.} {Comput.}, {Netw.}, {Storage} and {Anal.}}\hskip 1em plus 0.5em minus 0.4em\relax Denver CO USA: {ACM}, Nov. 2023, pp. 1--15.

\bibitem{zawishEnergyAI2023}
M.~Zawish, N.~Ashraf, R.~I. Ansari, and S.~Davy, ``Energy-aware {AI}-driven framework for edge-computing-based {IoT} applications,'' \emph{{IEEE} Internet of Things Journal}, vol.~10, no.~6, pp. 5013--5023, Mar. 2023.

\bibitem{3gpp_studyAImano_2023}
{3GPP}, ``Study on artificial intelligence / machine learning ({AI}/{ML}) management {(Release 18)},'' {3GPP}, {TR 28.908 v1.2.0}, Apr. 2023.

\bibitem{mat_brown_digging_2023}
\BIBentryALTinterwordspacing
{M}at {B}rown. (2023, May) Digging into data center efficiency, {PUE} and the impact of {HCI}. [Online]. Available: \url{https://www.nutanix.dev/2023/05/04/digging-into-data-center-efficiency-pue-and-the-impact-of-hci}
\BIBentrySTDinterwordspacing

\bibitem{ptflops}
\BIBentryALTinterwordspacing
V.~Sovrasov. (2023, Dec.) ptflops: A flops counting tool for neural networks in pytorch framework. [Online]. Available: \url{https://github.com/sovrasov/flops-counter.pytorch}
\BIBentrySTDinterwordspacing

\bibitem{zhang_EAIBench_2023}
F.~Zhang \emph{et~al.}, ``{EAIBench}: An energy efficiency benchmark for {AI} training,'' in \emph{14th {BenchCouncil} International Symposium}, A.~Gainaru, C.~Zhang, and C.~Luo, Eds., Nov. 2023, pp. 19--34.

\bibitem{anthony_carbontracker_2020}
L.~F.~W. Anthony, B.~Kanding, and R.~Selvan, ``Carbontracker: Tracking and predicting the carbon footprint of training deep learning models,'' Jul. 2020, arXiv:2007.03051 [cs, eess, stat].

\bibitem{lannelongue_green_2021}
L.~Lannelongue, J.~Grealey, and M.~Inouye, ``{Green Algorithms}: Quantifying the carbon footprint of computation,'' \emph{Advanced Science}, vol.~8, no.~12, May 2021.

\bibitem{oran_wg2_aiml_2021}
{O-RAN WG2}, ``{AI}/{ML} workflow description and requirements,'' {O-RAN} Alliance, Tech. Rep. 01.03, Jul. 2021.

\bibitem{rost_complexityrate_2015}
P.~Rost, S.~Talarico, and M.~C. Valenti, ``The complexity–rate tradeoff of centralized radio access networks,'' \emph{{IEEE} Trans. Wireless Commun.}, vol.~14, no.~11, pp. 6164--6176, Nov. 2015.

\bibitem{nikaein_processing_2015}
N.~Nikaein, ``Processing radio access network functions in the cloud: Critical issues and modeling,'' in \emph{Proc. of the 6th {Int.} {Workshop} on {Mobile} {Cloud} {Comput.} and {Services}}.\hskip 1em plus 0.5em minus 0.4em\relax Paris France: ACM, Sep. 2015, pp. 36--43.

\bibitem{ge_energy_2017}
X.~Ge, J.~Yang, H.~Gharavi, and Y.~Sun, ``Energy efficiency challenges of {5G} small cell networks,'' \emph{{IEEE} Commun. Mag.}, vol.~55, no.~5, pp. 184--191, May 2017.

\bibitem{landauer_irreversibility_1961}
R.~Landauer, ``Irreversibility and heat generation in the computing process,'' \emph{IBM J. Res. and Develop.}, vol.~5, no.~3, pp. 183--191, Jul. 1961.

\bibitem{khatibi_modelling_2018}
S.~Khatibi, K.~Shah, and M.~Roshdi, ``Modelling of computational resources for {5G} {RAN},'' in \emph{{European} {Conf.} on {Netw.} and {Commun.}}, Ljubljana, Slovenia, Jun. 2018, pp. 1--5.

\bibitem{castanoHugFaceCarbon2023}
J.~Castaño, S.~Martínez-Fernández, X.~Franch, and J.~Bogner, ``Exploring the carbon footprint of hugging face's {ML} models: A repository mining study,'' in \emph{{ACM}/{IEEE} International Symposium on Empirical Software Engineering and Measurement ({ESEM})}.\hskip 1em plus 0.5em minus 0.4em\relax {IEEE}, Oct. 2023, pp. 1--12.

\end{thebibliography}

\enlargethispage{-4.2in}

\begin{IEEEbiography}
[{\includegraphics[width=1in,height=1.25in,clip,keepaspectratio]{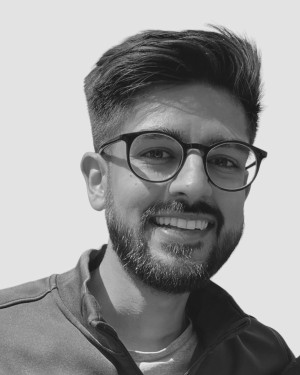}}]
{Kishan Sthankiya} received the H.E. certificate in biosciences from King's College London, U.K., in 2015. He was a professionally accredited Infrastructure Consultant with expertise across enterprise projects. He is pursuing an Eng.D. at the Data-Centric Engineering Centre for Doctoral Training
and Networks Research Group at Queen Mary University of London, since 2021. His research interests include next-generation radio access, sustainability and machine learning.
\end{IEEEbiography}
\vskip 0pt plus -1fil
\begin{IEEEbiography}
[{\includegraphics[width=1in,height=1.25in,clip,keepaspectratio]{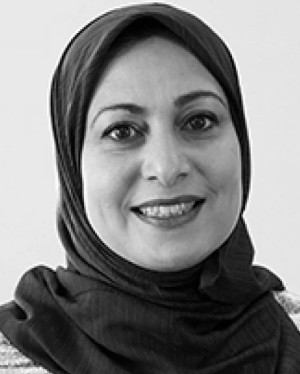}}]{Nagham Saeed} received the B.S degree in computer \& control and the M.S in mechatronics from the University of Technology, Baghdad, Iraq in 1992 and 1999, respectively, and the Ph.D. degree from Wireless Networks and Communications Centre, Brunel University, London, UK in 2011. Her Ph.D. research was optimizing Mobile Ad Hoc Wireless Communication Networks by introducing an Intelligent Mobile Ad Hoc Network System based on AI. Her research interests include expert systems for smart cities, wherein she applies AI algorithms to drive modeling and optimization. She is currently an Associate Professor in Electrical and Electronic Engineering and the Head of the Industrial Internet of Things Research Group at the University of West London. She is recognized as CEng by the Engineering Council, a senior member of the IEEE, a member of the IET, 2024-2025 IEEE UK \& Ireland Section Vice Chair, 2026-2027 IEEE UK \& Ireland Section Elect-Chair, IEEE Women in Engineering UK \& Ireland section Past Chair (2023). 
\end{IEEEbiography}
\vskip 0pt plus -1fil
\begin{IEEEbiography}
[{\includegraphics[width=1in,height=1.25in,clip,keepaspectratio]{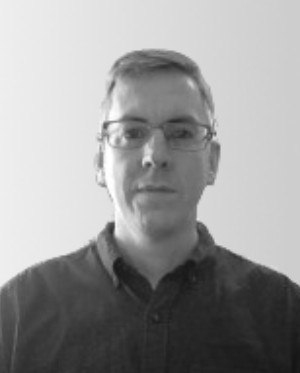}}]{Greg McSorley} received the B.A. degree in ancient history and archaeology from University of Manchester, U.K., in 2012 and the M.S. degree in geospatial information systems from Ulster University, Belfast, in 2022. His previous roles include management, archaeology and data analytics. He joined BT Research in 2022, and is currently an AI and Sustainability Researcher. His research interests include spatial analysis 
and net-zero initiatives.
\end{IEEEbiography}
\vskip 0pt plus -1fil
\begin{IEEEbiography}
[{\includegraphics[width=1in,height=1.25in,clip,keepaspectratio]{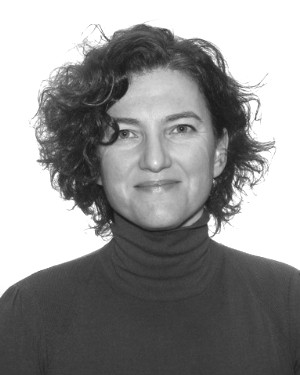}}]{Mona Jaber} received the B.E. degree in computer and communications engineering and the M.E. degree in electrical and computer engineering from the American University of Beirut, Lebanon, in 1996 and 2014, respectively, and the Ph.D. degree from the 5G Innovation Centre, University of Surrey, in 2017. Her Ph.D. research was on 5G backhaul innovations. She was a Telecommunication Consultant in various international firms with a focus on the radio design of cellular networks, including GSM, GPRS, 3G, and 4G. She led the IoT Research Group, at Fujitsu Laboratories of Europe, from 2017 to 2019, where she researched IoT-driven solutions for the automotive industry. She is currently a Lecturer in IoT with the School of Electronic Engineering and Computer Science, Queen Mary University of London. Her research interests include zero-touch networks, the intersection of ML and IoT in the context of sustainable development goals, and IoT-driven digital twins.
\end{IEEEbiography}
\vskip 0pt plus -1fil
\begin{IEEEbiography}
[{\includegraphics[width=1in,height=1.25in,clip,keepaspectratio]{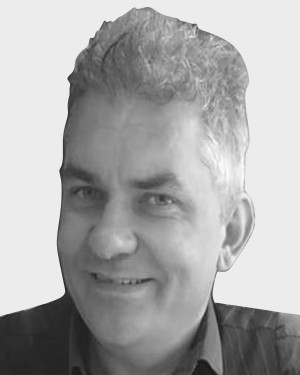}}]{Richard G. Clegg} is a Senior Lecturer in networks at the School of of Electronic Engineering and Computer Science, Queen Mary University of London. His research interests include complex networks and the statistics of network measurements. He is research lead at the company Pometry that develops software for temporal networks.
\end{IEEEbiography}

\EOD

\end{document}